\pdfoutput=1
\RequirePackage{fix-cm}
\documentclass[twocolumn,epjc3]{svjour3}  
\smartqed  
\RequirePackage{amsmath}
\RequirePackage{graphicx}
\RequirePackage{mathptmx}      
\RequirePackage{url}
\RequirePackage[colorlinks,citecolor=blue,urlcolor=blue,linkcolor=blue]{hyperref}
\journalname{Eur. Phys. J. C}
\begin{document}

\title{Comparing Traditional and Deep-Learning Techniques of Kinematic Reconstruction for Polarization Discrimination in Vector Boson Scattering}

\titlerunning{Comparing Reconstruction Techniques in VBS}        

\author{M. Grossi\thanksref{e1,pavia,ibm}
        \and
        J. Novak\thanksref{e2,jsi} 
         \and
        B. Ker\v{s}evan\thanksref{ul,jsi}
	 \and
	D. Rebuzzi\thanksref{pavia}
}

\thankstext{e1}{e-mail: michele.grossi01@universitadipavia.it}
\thankstext{e2}{e-mail: jakob.novak@ijs.si}


\institute{
Universit\'a di Pavia, Dipartimento di Fisica and INFN, Sezione di Pavia, %
Via A. Bassi 6, %
27100 Pavia, %
Italy\label{pavia}
\and
IBM Italia s.p.a. %
Circonvallazione Idroscalo, %
20090 Segrate (MI), %
Italy\label{ibm}
\and
Faculty of Mathematics and Physics, University of Ljubljana,
Jadranska cesta 19,
1000 Ljubljana,
Slovenia\label{ul}
\and
Jo\v{z}ef Stefan Institute,
Jamova cesta 39,
1000 Ljubljana,
Slovenia\label{jsi}
}

\date{Received: date / Accepted: date}

\maketitle

\begin{abstract}

Measuring longitudinally polarized vector boson scattering in $\mathrm{WW}$ channel is a promising way to investigate unitarity restoration with
 the Higgs mechanism and to search for possible physics beyond the Standard Model. In order to perform such a measurement, it is crucial to develop an efficient reconstruction of the full $\mathrm{W}$ boson kinematics in leptonic decays with the focus on polarization measurements. We investigated several approaches, from  traditional ones up to advanced deep neural network structures, and we compared their abilities in reconstructing the $\mathrm{W}$ boson reference frame and in consequently measuring the longitudinal fraction $\mathrm{W}_{\text{L}}$ in both semi-leptonic and fully-leptonic $\mathrm{WW}$ decay channels. 

\end{abstract}

\section{Introduction}
\label{sec:intro}

The Large Hadron Collider (LHC) at the European Organization for Nuclear Research (CERN) has, in the recent years, delivered unprecedented high-energy proton--proton collisions that have been collected and studied by the four LHC experiments, ALICE, ATLAS, CMS and LHCb. The  collaborations are exploiting this data to explore the fundamental nature of matter and the basic forces that 
shape the Universe, testing the Standard Model of particle physics (SM) in regimes never investigated before.
One process in particular, the Vector Boson Scattering (VBS), i.e.\@ the scattering of two massive vector bosons,
$\mathrm{VV} \to \mathrm{VV}$ with $\mathrm{V}$ = $\mathrm{W}$ or $\mathrm{Z}$, is the key for probing the electroweak sector of the SM in the TeV regime and the mechanism of 
electroweak symmetry breaking (EWSB)\cite{PhysRevLett.61.1053,PhysRevD.52.3878}. Experimentally, both validation of SM predictions and the discovery of new physics as unexpected deviations, are of great importance, in the short and medium terms. In the short term the studies are being performed by the experiments on the currently collected 
LHC data. In the medium term, LHC will be upgraded to the High Luminosity LHC project (HL-LHC) as an  European Strategy Forum on Research Infrastructures (ESFRI) landmark\footnote{http://roadmap2018.esfri.eu}, therefore providing orders of magnitude larger statistics in the future.

The VBS measurement is extremely challenging, because of its low signal yields, complex final states and large backgrounds. Its understanding requires a 
coordinated effort of experimental and theoretical physicists, with a wide spectrum of skills: detector knowledge, event reconstruction and 
simulation, and data mining. Only an interdisciplinary approach, as is currently indeed taking place, allows the best exploitation of the LHC 
data. 

In more detail, the VBS production involves quartic gauge boson self interactions, and the $s-$ and $t-$channel exchanges of a gauge boson or of a 
Higgs boson. In the SM, the contribution of the Higgs boson (H), discovered at the LHC \cite{HIGG-2012-27,CMS-HIG-12-028}, regularizes the VBS  amplitude by canceling divergences arising from longitudinally polarized vector bosons at high energy, as an explicit consequence of the EWSB \cite{PhysRevLett.38.883,PhysRevD.93.015004}. Furthermore,
 any enhancements in the VBS cross section beyond the SM prediction, as studied in  many scenarios investigating  physics beyond the SM, could
  potentially be detected at the LHC \cite{PhysRevD.87.093005,PhysRevD.87.055017}. 
  
  At the LHC, the dominant VBS processes result in final states with two gauge bosons and two jets
   ($\mathrm{VV} j j$). The same final state can however be achieved also through processes that do not involve VBS but are a result of other 
  electroweak-mediated and/or QCD-mediated contributions. Studies have shown that the same-sign $\mathrm{W}^\pm \mathrm{W}^\pm j j $ production has the largest VBS 
  contribution to the production cross section \cite{szleper2014higgs}, because tree-level Feynman diagrams not involving the self interactions are absent in the $s-$channel
   and suppressed in other channels. Consequently, the same-sign $\mathrm{W}^\pm \mathrm{W}^\pm j j $ production is well suited for EWSB and new physics 
   studies involving VBS at the LHC. The observation of the $\mathrm{W}^\pm \mathrm{W}^\pm jj$ electroweak production was already reported by the 
   ATLAS \cite{STDM-2017-06} and CMS \cite{CMS-SMP-17-004} collaborations using data recorded in Run 2 at a center-of-mass energy of $\sqrt{s}$ = 13 TeV.

The main aim of the study presented in this paper is to find an optimal method to identify the contribution of longitudinally polarized $\mathrm{W}$ 
bosons in the VBS process, by investigating strategies to reconstruct the event kinematics and thereby develop a technique 
that would efficiently discriminate between the longitudinal contribution and the rest of the participating processes in the VBS scattering. Several approaches have been studied and are presented in the following. The results demonstrate the advantages of machine-learning (ML) based techniques, 
which profit from optimal phenomenological observables in several stages of the procedure. The studies presented
have been performed on a VBS prototype analysis, to demonstrate the validity of the approach, but the same strategies can be implemented in any ATLAS and CMS analysis which involves VBS and/or $\mathrm{W}$ bosons.

\section{Vector Boson Scattering kinematics and polarization properties}
\label{sec:kin}

\begin{figure}[h]
\centering
\includegraphics[width=0.3\textwidth]{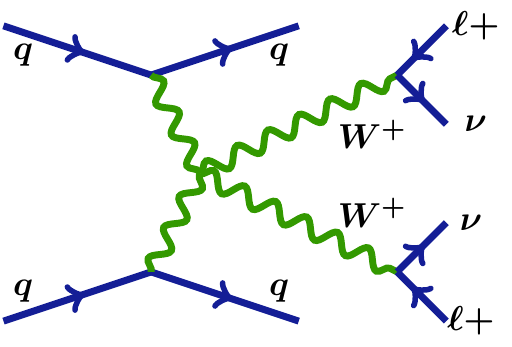}
\caption{A representative tree-level Feynman diagram of the VBS process at the LHC, involving a quartic weak boson coupling. The scattered quarks hadronize to jets and the $\mathrm{W}$ bosons decay leptonically in the given example, resulting in two (undetectable) neutrinos in the final state.}
\label{fig:VBS_diag}
\end{figure}

A representative Feynman diagram of the $\mathrm{W^\pm W^\pm} j j$ VBS process at the LHC is shown in Fig.~\ref{fig:VBS_diag}, where the gauge bosons are radiated off the incoming quarks and then scattered via the quartic self interaction vertex.
In this process, the two jets, produced by the fragmentation of the scattering partons, are expected to fly in predominantly 
forward direction of the opposite hemispheres, i.e. with large rapidity difference, while the $\mathrm{W}$ boson products are expected in the central region. This is a typical signature for this type of a scattering process \cite{Ballestrero:2018}. 
The two $\mathrm{W}$ bosons can decay either leptonically or hadronically, whereby leptonic decays into light leptons (electrons and muons) are experimentally preferred due to a lower background contamination and more efficient data acquisition triggers. In particular, the first observation papers of the ATLAS and CMS collaborations, cited above, required same-sign charged di-lepton pairs in the final state. In this paper, we
evaluate two scenarios, the semi-leptonic case where one of the $\mathrm{W}$ bosons decays into light leptons and the fully-leptonic case where both 
$\mathrm{W}$ bosons decay leptonically.

The EW scattering of the two $\mathrm{W}$ bosons has no strong kinematic constraints on the $\mathrm{W^\pm W^\pm}$ 
rest frame because it is a non-resonant process - the opposite would be true if the $\mathrm{W^\pm W^\pm}$ pair would be produced by a heavy resonance decay. Consequently, 
the constraints on the event kinematics are less stringent, which, as we will show in the following, makes the full  
reconstruction from the experimentally measurable kinematic quantities (i.e.\@ angles and charged particle four-momenta) much 
more difficult in the presence of non-measurable (one or two) neutrinos being produced in decay of the $\mathrm{W^\pm W^\pm}$ pair. 

Full kinematic reconstruction of the event is essential in view of being able to efficiently disentangle the different polarization contributions of the $\mathrm{W}$ bosons in the VBS process. The polarization fractions of the $\mathrm{W}$ boson are reflected directly in the angular distributions of the resulting leptons.
In absence of kinematic cuts on the leptons, the relation is very straightforward \cite{Ballestrero:2017bxn}, as one can write the angular distribution in the $\mathrm{W}$ boson rest frame as follows:
\begin{eqnarray}
\frac{1}{\sigma} \frac{\mathrm{d} \sigma}{\mathrm{d} \cos \vartheta}\left(\mathrm{W}^{\pm} \rightarrow \ell^{\pm} \nu\right) &=&\frac{3}{4} f_{0} \sin ^{2} \vartheta+ \frac{3}{8} f_{\text{R}}(1 \pm \cos \vartheta)^{2} \notag \\ &+&\frac{3}{8} f_{\text{L}}(1 \mp \cos \vartheta)^{2},
\end{eqnarray}
where the polarization fractions $\sum_{i=0,\text{L,R}} f_i =1$ weigh the Legendre polynomials of the lepton polar angle  $\vartheta$, defined as the angle between the lepton in the $\mathrm{W}$ rest frame with respect to the $\mathrm{W}$ direction in the laboratory frame.

As argued in the papers Ref.~\cite{Ballestrero:2017bxn,Ballestrero:2019qoy}, this relation becomes more complex in the presence of kinematic cuts as well as reconstruction inefficiencies and ambiguities. The principal challenge is thus being able to reconstruct the lep\-ton\-ically-decaying $\mathrm{W}$ boson rest frame(s), i.e.\@ to fully reconstruct the neutrino momenta. 

In what follows, we will present several approaches for trying to achieve this goal, in 
both semi-leptonic and fully-leptonic VBS cases, where we make use of simple cuts, advanced kinematic variables and machine learning approaches.

\section{Simulated sample definition}
\label{sec:sim}
VBS events are generated using \texttt{PHANTOM} \cite{Ballestrero:2007xq}, a tree level Monte Carlo generator for six parton final states at
 $O(\alpha_{\text{EW}}^6)$ and $O(\alpha_{\text{EW}}^4 \cdot \alpha_s^2)$ in perturbation theory, including possible interferences between the two sets of diagrams (see  \ref{sec:app1} for more details on the generator and \ref{sec:app2} for the event generation configuration).
In order to study the impact of experimental measurements, generated events are processed with \texttt{Delphes} \cite{Delphes:2014}, a framework for fast simulation of collider experiments, which mimics the detector 
effects on the measured quantities, such as final state physics objects. This fast simulation includes a generic tracking system, embedded into a magnetic field, 
electromagnetic and hadronic calorimeters and a muon system. Each of the subsystems can be customized to reproduce the effect of any specific experiment sub-detector.
 In the present study, the ATLAS configuration (an approximate ATLAS detector description) has been adopted\footnote{\url{https://cp3.irmp.ucl.ac.be/projects/delphes/browser/git/cards/delphes_card_ATLAS.tcl}}.  These studies show that our findings at the truth (generator-level) level and the (approximate) \texttt{Delphes} simulation are qualitatively the same: since we use only measurable quantities in our studies, only a degraded resolution on these quantities, introduced by experimental uncertainties, is added by detector simulation.

\section{Reconstruction of the $\mathrm{W}$ Boson reference frame using kinematic cuts}
\label{sec:rec}

\subsection{$\mathrm{W}$ boson decay kinematics}
\label{subsec:wkine}

In this paper we aim at systematically evaluating strategies to reconstruct the angular distribution of leptons in the $\mathrm{W}$ rest frame. The success of this reconstruction consequently facilitates the extraction of $\mathrm{W}$ boson polarization fractions.
In order to reconstruct the $\mathrm{W}$ boson rest frame, one needs to reconstruct the full event kinematics, including  momenta of  neutrinos resulting from $\mathrm{W}$ boson decays. In case there is only one neutrino in the final state,  the transverse components of the neutrino momentum can be extracted by measuring the missing transverse momentum (MET). This happens in the semi-leptonic VBS channel, when one $\mathrm{W}$ boson decays into light leptons (electrons and muons) and the other $\mathrm{W}$ boson decays hadronically.
In the fully-leptonic channel, where both $\mathrm{W}$ bosons decay into light leptons, resulting in two neutrinos in the final state, both neutrinos contribute to the measured MET and consequently the transverse momenta of the two neutrinos cannot be determined directly. The longitudinal neutrino momentum cannot be determined in either case due to the unknown center-of-mass system of the colliding partons in a proton-proton collision. 

In the semi-leptonic case, we can obtain the longitudinal momentum component of the single neutrino by introducing the
 $\mathrm{W}$ boson mass constraint and solve if for the longitudinal neutrino component, $p_{\nu\text{L}}$. 
To this end, let us first consider a single $\mathrm{W}$ boson decaying leptonically in the laboratory frame. Requiring the four-momentum conservation of the $\mathrm{W}$ boson decay products in the ultra-relativistic limit, the second-order equation can be easily obtained:
\begin{eqnarray}
&\underbrace{\left(p^2_{\ell \text{L}} -E_\ell ^2\right)}_{a} p_{\nu \text{L}}^2 \notag \\
&+ \underbrace{\left( { m_{\mathrm{W}}^2} p_{\ell \text{L}} +2 p_{\ell \text{L}} (\vec{p}_{\ell \text{T}} \cdot 
\vec{p}_{\nu \text{T}})\right)}_{b} p_{\nu \text{L}} \notag \\
&+ \underbrace{\frac{m_{\mathrm{W}}^4}{4}+ ( \vec{p}_{\ell \text{T}}\cdot\vec{p}_{\nu \text{T}})^2 + {m_{\mathrm{W}}^2} (\vec{p}_{\ell \text{T}}\cdot \vec{p}_{\nu \text{T}}) - E_\ell ^2 \vec{p}_{\nu \text{T}}^ {\;2} }_{c} = 0
\label{eq:soeq}
\end{eqnarray}
where $m_{\mathrm{W}}$ is the $\mathrm{W}$ boson invariant mass, $p_{\ell \text{L}}$, $\vec{p}_{\ell \text{T}}$ are respectively the longitudinal and transverse components of the lepton momentum and $E_\ell$ represents its energy, while $\vec{p}_{\nu \text{T}}$ denotes the transverse neutrino momentum component, which we can equate with MET.

Eq.~\ref{eq:soeq} defines the unknown $p_{\nu \text{L}}$ as the zeros of a second-order polynomial, therefore the problem of a negative discriminant ($\Delta = b^2 - 4ac < 0$) should also be considered. Some {\em ad-hoc} solutions have been adopted by the experimental analysis groups, such as setting the discriminant to zero or recalculating the discriminant by applying a $\mathrm{W}$ transverse mass constraint. In this context,  speculations on the meaning of the imaginary solutions are also considered.
In the present study, we ignore the negative discriminant cases, while we focus only on the ambiguity of sign ($+/-$) in the two solutions for the positive discriminant  case. There is a priori no physical reason to prefer one solution instead of the other.

In the following, we will comment on how this ambiguity could be addressed and resolved, by adopting specific physics selection criteria.

\subsection{Semi-leptonic VBS channel}
\label{subsec:semil}

\begin{figure}
	\centering
	\includegraphics[width=0.45\textwidth]{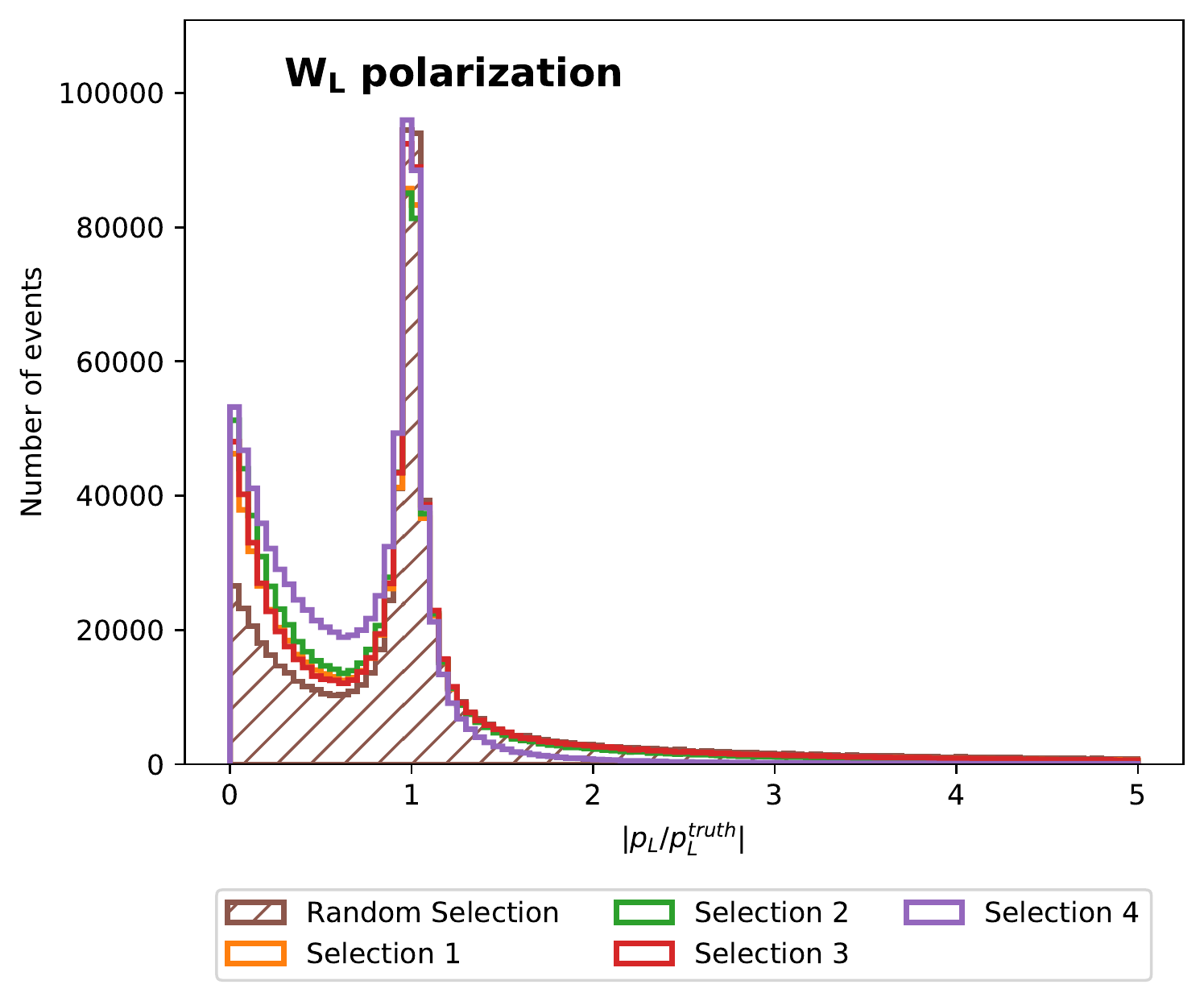}
	\includegraphics[width=0.45\textwidth]{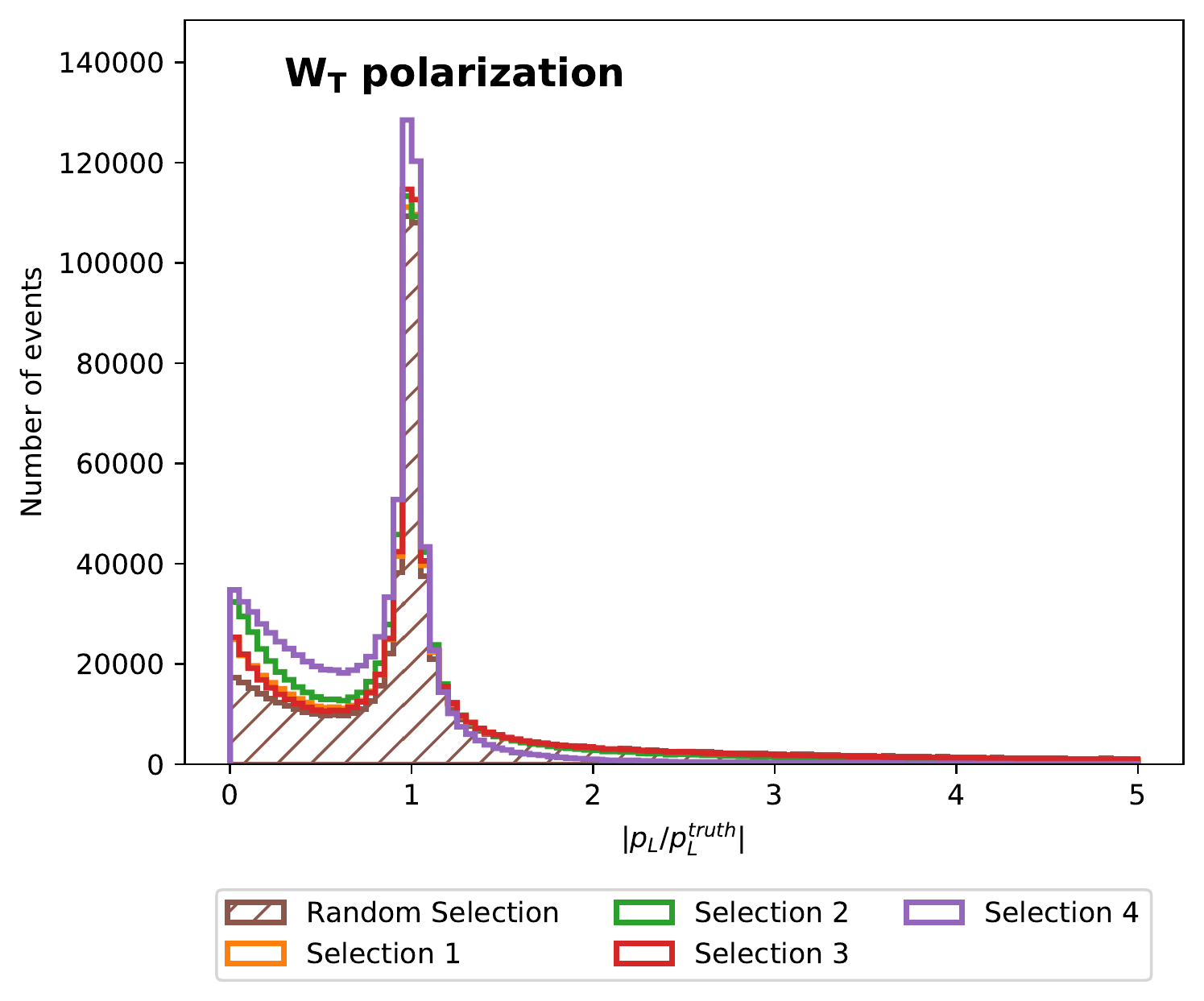}
	\caption{Reconstruction of longitudinal neutrino momentum using different selection criteria described in the text. The random selection is shown as the worst-case scenario. 
The ratio between the longitudinal neutrino momentum $p_{\nu \text{L}}$ and the true (generator-level) value $p_{\nu \text{L}}^\mathrm{truth}$ is shown. The top plot shows results obtained assuming the purely longitudinal ($\mathrm{W}_{\text{L}}$) and the bottom plot the purely transverse ($\mathrm{W}_{\text{T}}$) scenario.}
\label{fig:plreco}
\end{figure}

\begin{figure}
	\centering
	\includegraphics[width=0.45\textwidth]{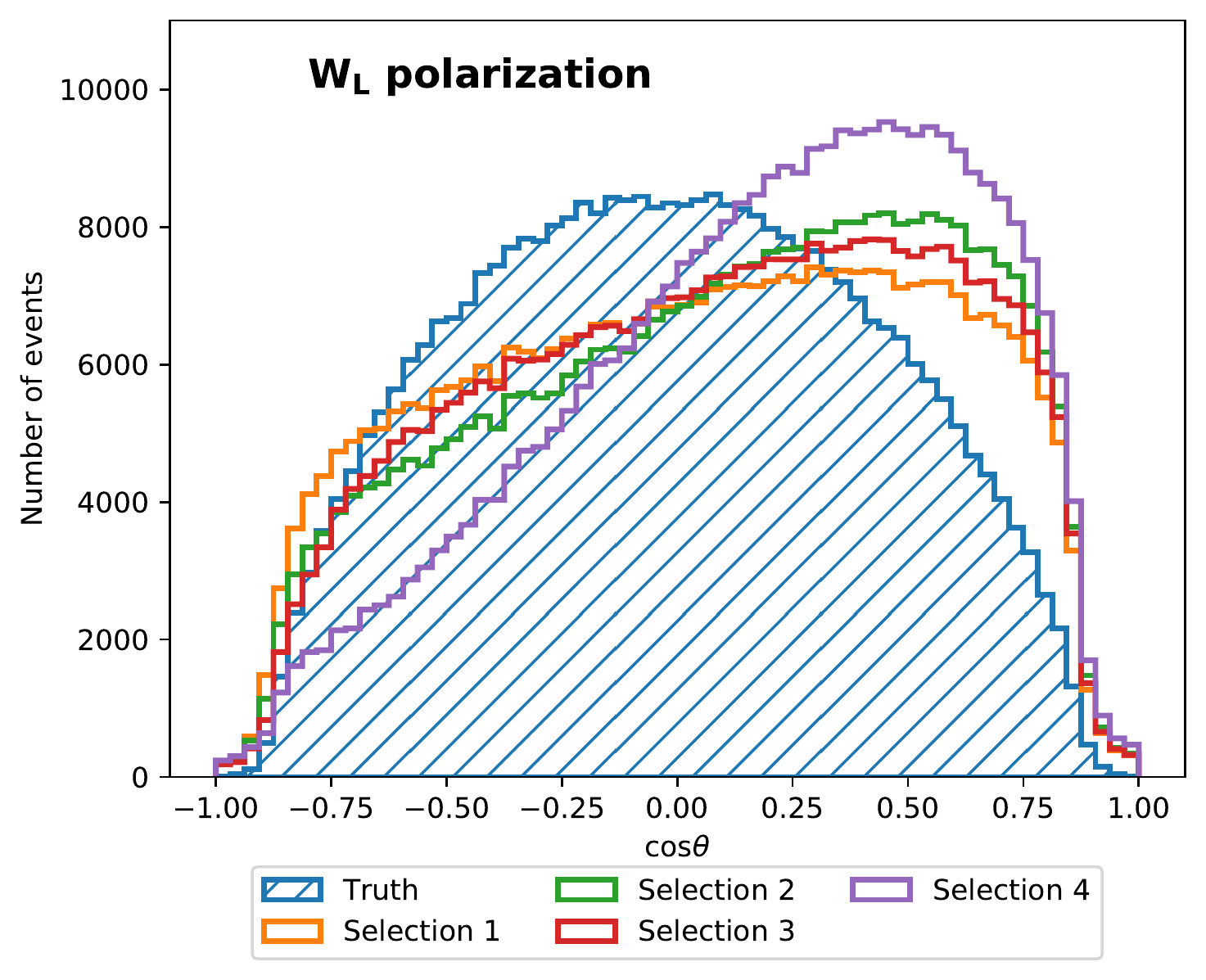}
	\includegraphics[width=0.45\textwidth]{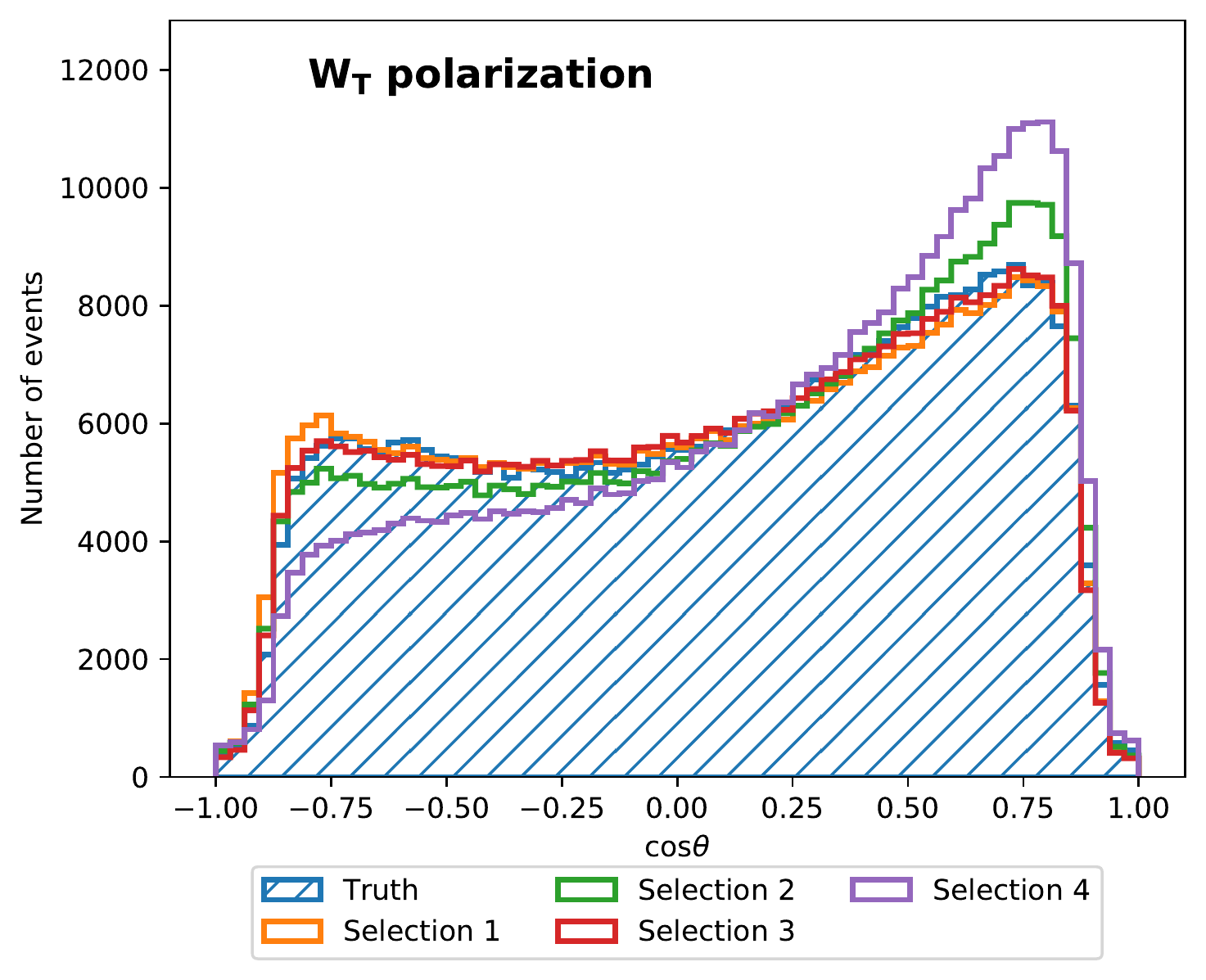}
	\caption{Shape of the reconstructed lepton angular distribution in the reconstructed $\mathrm{W}$ boson rest frame. 
Parton-level (truth) distributions are shown for comparison, using the generator-level (true) $\cos\vartheta$.
	The top plot shows  results obtained assuming the purely longitudinal ($\mathrm{W}_{\text{L}}$) and the bottom plot the purely transverse ($\mathrm{W}_{\text{T}}$) scenario. 
	Selection criteria 1--4, have been studied as strategies for choosing the correct solution, as described in the text.}
	\label{fig:seltheta}
\end{figure}

The VBS semi-leptonic channel consists of a hadronically-decaying $\mathrm{W}$ boson with a large
transverse momentum $p_{\text{T}}$, a $\mathrm{W}$ boson decaying to a light charged lepton (electron or
muon) and a neutrino, with additional jets in the forward regions of the detector. This signature benefits from 
a larger branching ratio with respect to the fully-leptonic one,
thanks to the hadronic decay of one vector boson, but it faces a higher reducible
background from $\mathrm{W}$+jets production and an intrinsic ambiguity at detector level
in identifying the jets coming from the $\mathrm{W}$ boson decay.
As discussed in the previous Subsection~\ref{subsec:wkine}, the longitudinal neutrino momentum from the leptonically decaying $\mathrm{W}$ boson can be reconstructed  by using the  measured  MET together with the lepton four-momentum, and applying the $\mathrm{W}$ mass constraint. The remaining ambiguity is the choice of the sign in Eq.~\ref{eq:soeq}.

\begin{figure*}
	\includegraphics[scale=0.55]{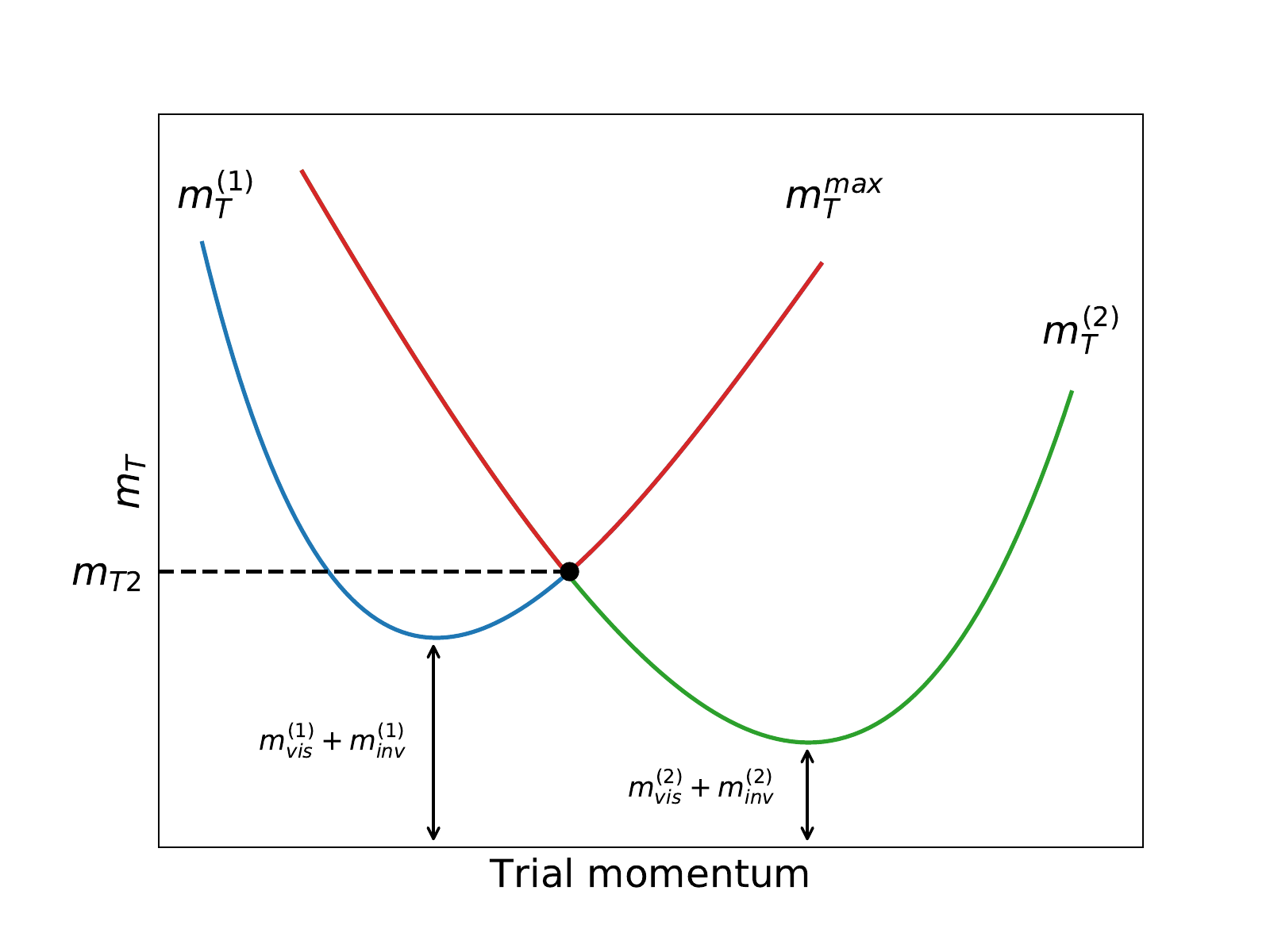}
	\includegraphics[scale=0.55]{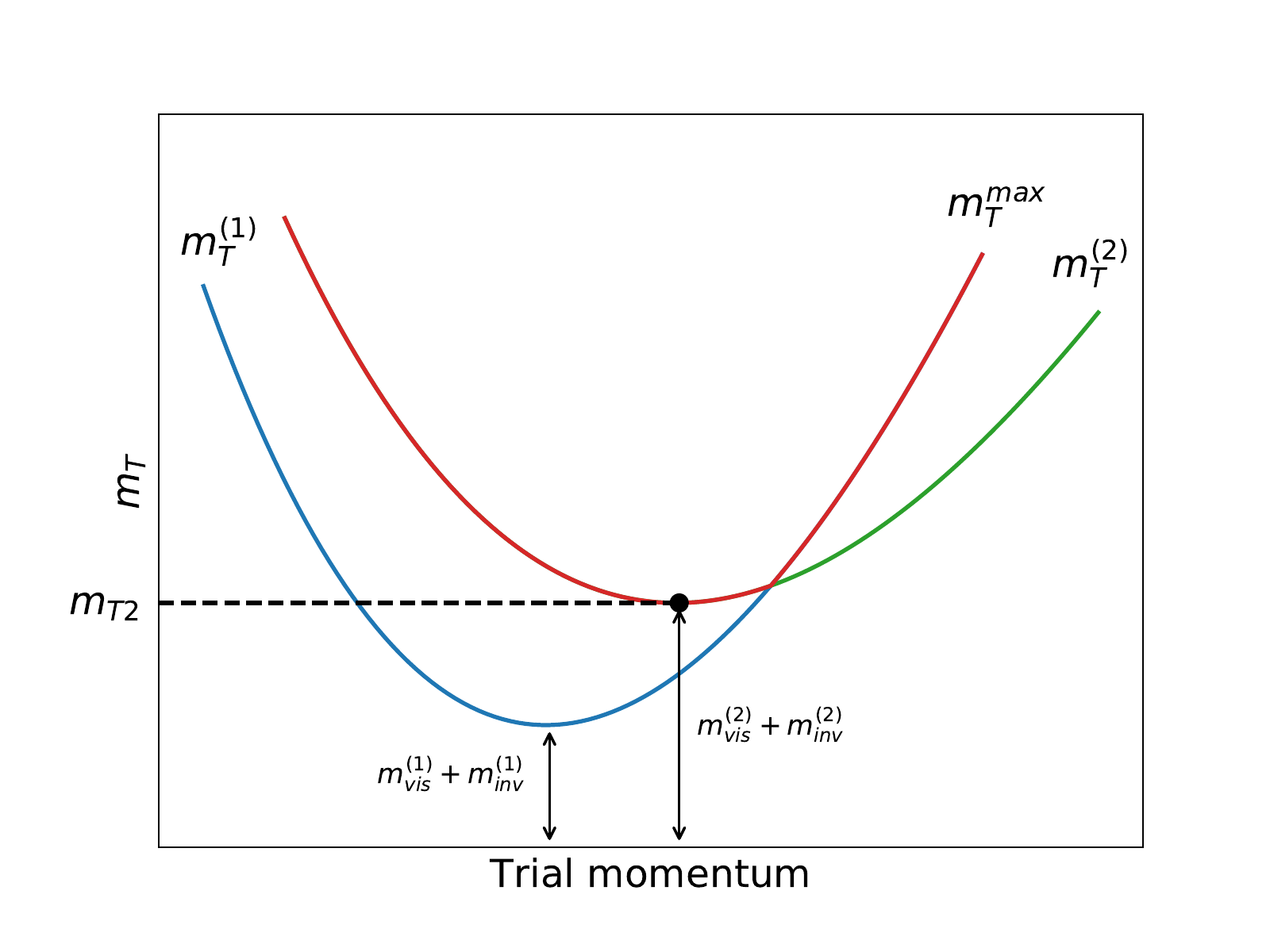}
	\caption{The balanced and unbalanced solution classes of the MAOS minimization problem. The two  transverse masses $m_{\text{T}}^{(1,2)}$ as a function of the trial momenta are shown in blue and green and the resulting functional $m_{\text{T}}^{\text{max}}$ is shown in red. In the balanced case (left) the solution $m_{\text{T2}}$ is in the intersection of the two $m_{\text{T}}$ curves and in the unbalanced case (right) the $m_{\text{T2}}$ is in one of the minima of the $m_{\text{T}}$ curves. The  $m^{(1,2)}_{\text{vis}}$ and $m^{(1,2)}_{\text{inv}}$, represent the masses of the visible and invisible final state particles. In the case studied in this paper, where the masses of the final state particles are negligible in comparison to the mass of the decaying heavy particle, the $m^{(1)}_{\text{T}}$ and $m^{(2)}_{\text{T}}$ have their minima values approximately at zero and therefore only the balanced solution topology is possible.}
	\label{fig:maossolution}       
\end{figure*}

Our studies have determined that this sign ambiguity can be partially resolved by adopting certain decision criteria. The criteria are a result of extensive studies of the kinematics in the VBS simulation and show that the correct solution lies in a specific kinematic region, which we, at this stage, try to define with specific kinematic cuts on reconstructed quantities. 
That being said, our studies show this kinematic region exists, but the demarcation of this kinematic region is not straightforward. The application of machine learning approaches to identify its boundaries seems a promising approach. 

The cut-based decision criteria require the correct solution to pass a certain kinematic cut, if both solutions (i.e.\ using either sign) pass or fail this kinematic cut, a random solution is chosen in our evaluations. Selection criteria considered are the following ($a$ and $b$ represent the parameters of the quadratic Eq.~\ref{eq:soeq}):

\begin{itemize}
\item Random Selection: sign is chosen randomly (this is the worst-case scenario);
\item Selection $1$: the correct solution is required to have the absolute value of the scalar product of the reconstructed neutrino three-momentum with the reconstructed $\mathrm{W}$ three-momentum smaller than 5000 GeV$^2$, random solution is chosen if both solutions pass/fail this criterion;
\item Selection $2$: the correct solution is required to have the absolute value $|p_{\nu \text{L}}|$  smaller than 50 GeV, random solution is chosen if both solutions pass/fail this criterion;
\item Selection $3$: the correct solution is required to have the absolute value of the scalar product of the reconstructed neutrino three-momentum with the reconstructed $\mathrm{W}$ three-momentum, multiplied by $a/b$, smaller than 25 GeV, random solution is chosen if both solutions pass/fail this criterion;
\item Selection $4$: the correct solution is required to lie above the parabola minimum, i.e. $p_{\nu \text{L}} > -b/(2a)$ or, alternatively, in a normalized way: $p_{\nu \text{L}} \cdot a/b > -0.5$, random solution is chosen if both solutions pass/fail this criterion.
\end{itemize}

The effect of the different selection criteria on the reconstruction of the longitudinal neutrino momentum $p_{\nu \text{L}}$
from generator-level measurable quantities can be appreciated in Fig.~\ref{fig:plreco}, which shows the ratio between this quantity and the true (generator-level) value $p_{\nu \text{L}}^\mathrm{truth}$, using simulated samples implementing the scenarios where the leptonically decaying $\mathrm{W}$ boson is longitudinally  ($\mathrm{W}_{\text{L}}$) and transversely ($\mathrm{W}_{\text{T}}$) polarized. 

A qualitative analysis of this plot demonstrates that all selections  enhance the peak at one, as we aim to achieve, whereby selection 1 coincides with (lies under) selection 3. As a distinctive case, the selection 4 achieves a higher peak at one by substantially reducing the high-value tails, while the other selections reduce the contributions which are less than one. 
In general the widths are limited from below by the resolution on the reconstructed $\mathrm{W}$ momentum.

Comparisons of truth and reconstructed lepton angular distribution in the $\mathrm{W}$ boson reference frame using generator-level measurable quantities from the same samples are shown in Fig.~\ref{fig:seltheta}. 
A quantitative metric to asses the performance of the selection criteria is the fraction of correct solution choices, where we define as correct a solution giving $p_{\nu \text{L}}$ with a smaller absolute error with respect to the true value. The results for different polarizations of the leptonically decaying $\mathrm{W}$ boson ($\mathrm{W}_{\text{L}}$, $\mathrm{W}_{\text{T}}$), unpolarized $\mathrm{W}$ boson generated in the OSP framework\footnote{OSP refers to a method called On Shell Projection (OSP), an approximation implemented by {\tt PHANTOM} to compute the amplitude of resonant contributions projecting (in the numerator) the four momenta of the decay particles on shell (see \ref{sec:app1} for details).}, and full computation (unpolarized $\mathrm{W}$ boson, with non-resonant production modes included) are reported in Table~\ref{tab:solfrac}.

\begin{center}
\begin{table}
	\caption{Summary of the fractions of correct choices in the semi-leptonic channel for different selection criteria, applied on samples containing a purely longitudinally polarized ($\mathrm{W}_{\text{L}}$) leptonically decaying $\mathrm{W}$ boson, a purely transversely polarized ($\mathrm{W}_{\text{T}}$) boson, an unpolarized $\mathrm{W}$ boson generated in the OSP {\tt PHANTOM} framework (Un. OSP) and the full computation with an unpolarized $\mathrm{W}$ boson and with non-resonant production modes included (Full comp.) is shown.  }
	\label{tab:solfrac}       
	\begin{tabular}{llllll}
		\hline\noalign{\smallskip}
		Type & $\mathrm{W}_{\text{L}}$ & $\mathrm{W}_{\text{T}}$ & Un. OSP & Full comp. \\
		\noalign{\smallskip}\hline\noalign{\smallskip}
		Selection 1 & 0.503 & 0.544 & 0.527 & 0.551\\
		Selection 2 & 0.518 & 0.580 & 0.554 & 0.576\\
		Selection 3 & 0.528 & 0.553 & 0.539 & 0.564\\
		Selection 4 & 0.565 & 0.645 & 0.607 & 0.638\\
		\noalign{\smallskip}\hline
	\end{tabular}
\end{table}
\end{center}

To summarize, in the semi-leptonic channel, the reconstruction of the $\mathrm{W}$ reference frame can be obtained up to a neutrino sign ambiguity. 
The use of selection criteria presented here to 
select the correct solution is not fully effective and could be improved by using machine learning techniques, as described in Section~\ref{sec:dnn}.

\subsection{Fully-leptonic VBS channel}
\label{sec:fullep}

Here we extend and adapt the approach described in the previous section to the fully-leptonic VBS case, which is the most promising process in terms of background discrimination, despite its lower cross section due to $\mathrm{W}$ boson decay branching ratios. 
In this case, the kinematic is more complex, with two neutrinos in the final state, and the reconstruction of the two $\mathrm{W}$ boson rest frames represents a real challenge from an experimental point of view.
There are eight unknown parameters and only six equations defining kinematic constraints\footnote{The eight kinematic unknowns are the two four-momenta ($E,p_X,p_Y,p_Z$) for each neutrino and the six kinematic constraints are two mass constraints (on-shell requirements) for the two neutrinos, two mass constraints for the two $\mathrm{W}$ bosons and the two sums of  momenta of the two neutrinos in transverse directions (x,y) resulting in MET components.}. 

Several algorithms were developed to fully reconstruct the kinematics of a process with two invisible particles in the final state. The 
$\texttt{MT2}$-Assisted On-Shell (MAOS) \cite{Cho:2007dh} algorithm we applied in this study exploits the fact that most of the processes involving an exchange of the $\mathrm{W}$ boson (or another heavy decaying particle) prefer the phase space region with a low  transverse mass of the $\mathrm{W}$ boson.
This approach has been successfully used in SUSY searches and also in $\mathrm{H \rightarrow WW}$ reconstruction \cite{Sonnenschein:2006ud,Cho:2007dh,Choi:2009hn}.

In order to simplify the notation, let us restrict ourselves to a VBS process where one $\mathrm{W}$ decays to a muon and the other to an electron. The MAOS algorithm introduces a functional expression of trial (unknown) neutrino transverse momenta $\vec{p}_1$ and $\vec{p}_2$ as: 
\begin{equation}
m_{\text{T}}^{\text{max}}(\vec{p}_1,\vec{p}_2) = \max[m^{(1)}_{\text{T}}(\vec{p}_1),m^{(2)}_{\text{T}}(\vec{p}_2)],
\end{equation}
 with:
\begin{eqnarray}
m^{(1)}_{\text{T}}(\vec{p}_1) &= 2(|\vec{p}_{\mathrm{T}}^{\,\mu}||\vec{p}_1| - \vec{p}_{\mathrm{T}}^{\,\mu}\cdot\vec{p}_1),\notag \\
m^{(2)}_{\text{T}}(\vec{p}_1)&= 2(|\vec{p}_{\mathrm{T}}^{\,e}||\vec{p}_2| - \vec{p}_{\mathrm{T}}^{\,e}\cdot\vec{p}_2),
\end{eqnarray}
\noindent
where $\vec{p}_{\mathrm{T}}^{\,\mu}$ and $\vec{p}_{\mathrm{T}}^{\,e}$ represent muon and electron transverse momentum respectively. The minimum of this functional defines the $m_{\text{T}2}$ variable and represents a compromise for  minimization of both $\mathrm{W}$ transverse masses at the same time:
\begin{equation}
m_{\text{T2}} \equiv \min_{\vec{p}_1 + \vec{p}_2 = \vec{p}_{\text{T}}^{\,\text{miss}}} m_{\text{T}}^{\text{max}}(\vec{p}_1, \vec{p}_2) = \left. m_{\text{T}}^{\text{max}}\right|_{\vec{p}^{\,\nu_{\mu}}_{\text{T}},\vec{p}^{\,\nu_e}_{\text{T}}},
\label{Eq:MAOS}
\end{equation}
with the solutions $\vec{p}^{\,\nu_e}_{\text{T}}$ and $\vec{p}^{\,\nu_{\mu}}_{\text{T}}$ giving the estimates for the neutrino transverse momenta.

\begin{figure}
\centering
\includegraphics[width=.45\textwidth]{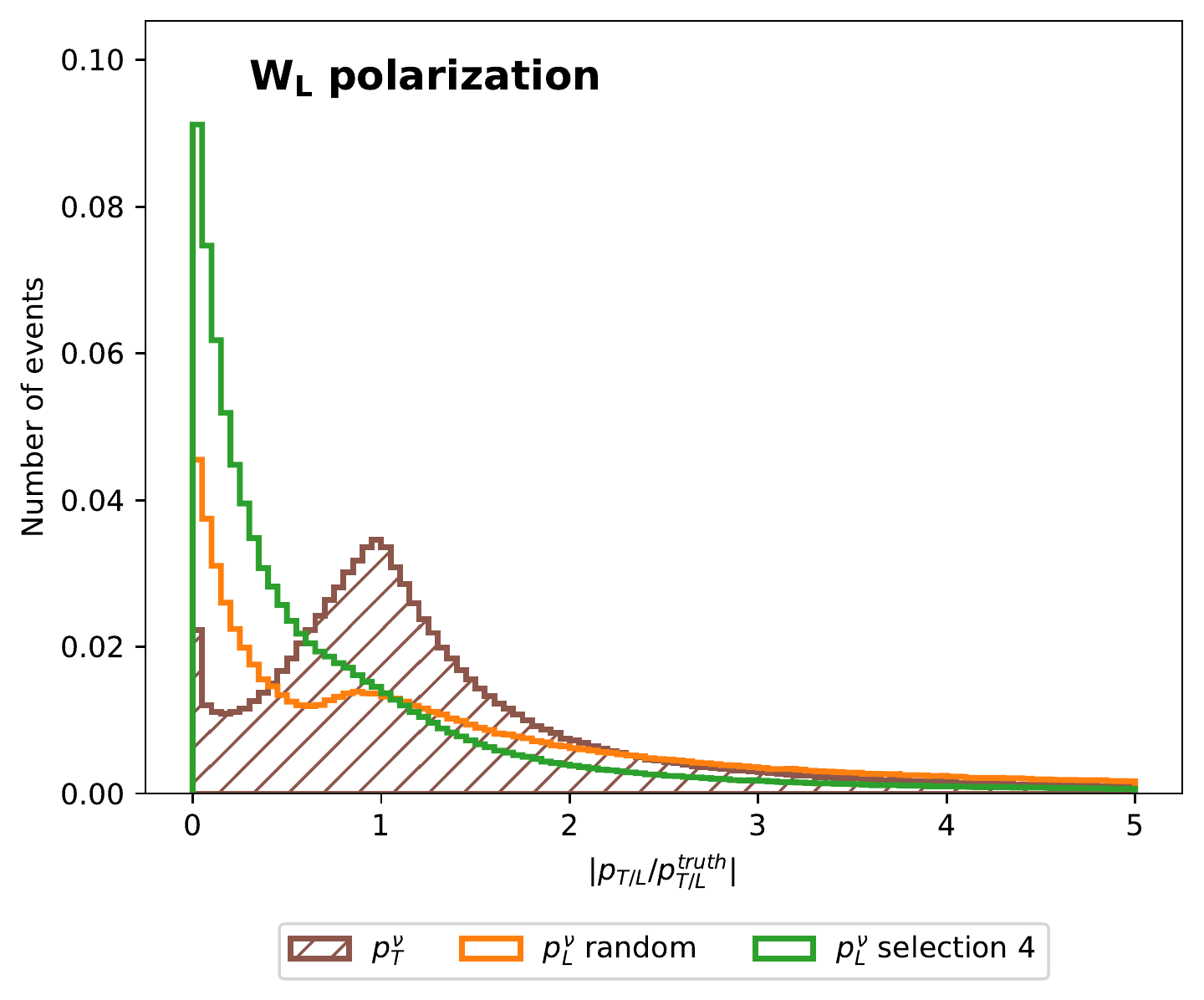}
\includegraphics[width=.45\textwidth]{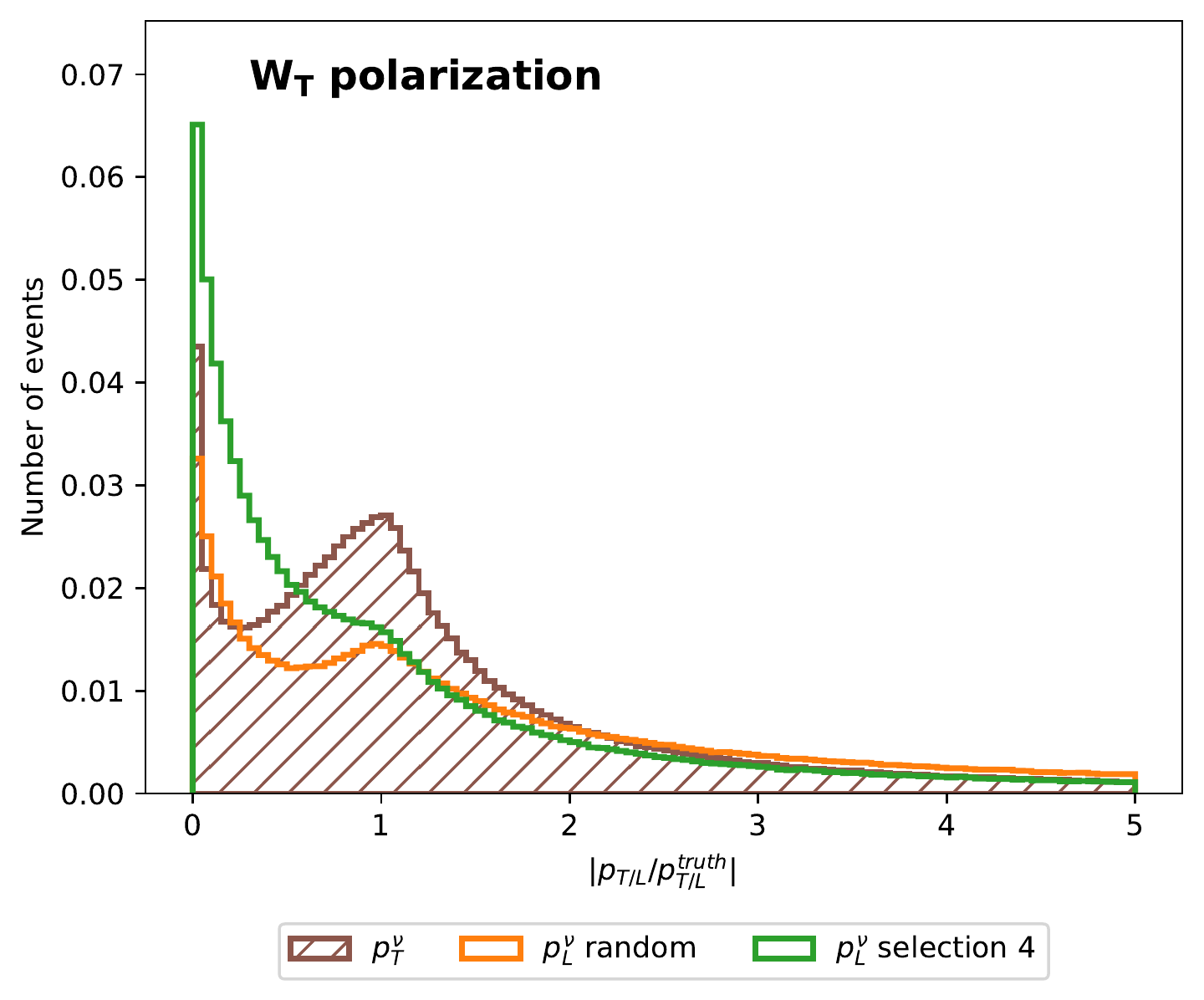}\\
\caption{Reconstruction of the neutrino momenta in the fully-leptonic channel at the truth (generator) level using the MAOS approach. The ratio between the reconstructed and true neutrino momentum is shown for the longitudinal and transverse components. 
The top plot shows results obtained assuming the purely longitudinal ($\mathrm{W}_{\text{L}}$) and the bottom plot the purely transverse ($\mathrm{W}_{\text{T}}$) scenario. 
While the transverse momenta are reconstructed using the MAOS procedure, the $\mathrm{W}$ mass constraint from the semi-leptonic case is adopted for reconstruction of both longitudinal neutrino momenta. The random choice and selection 4, as described in Section~\ref{subsec:semil}, were studied as strategies to resolve solution ambiguities.}
\label{fig:pmaos}
\end{figure}

\begin{figure}
\centering
\includegraphics[width=0.45\textwidth]{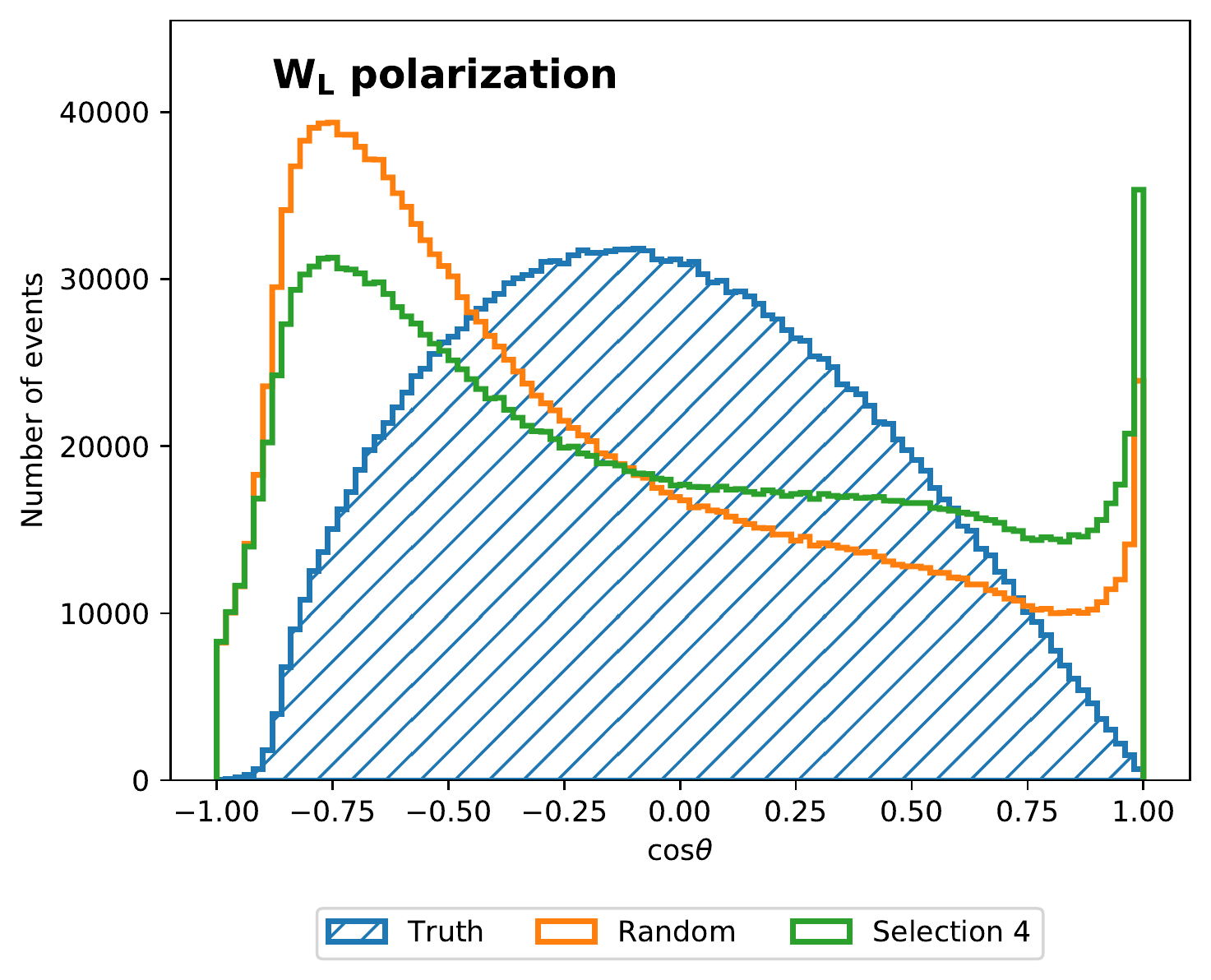}	
\includegraphics[width=0.45\textwidth]{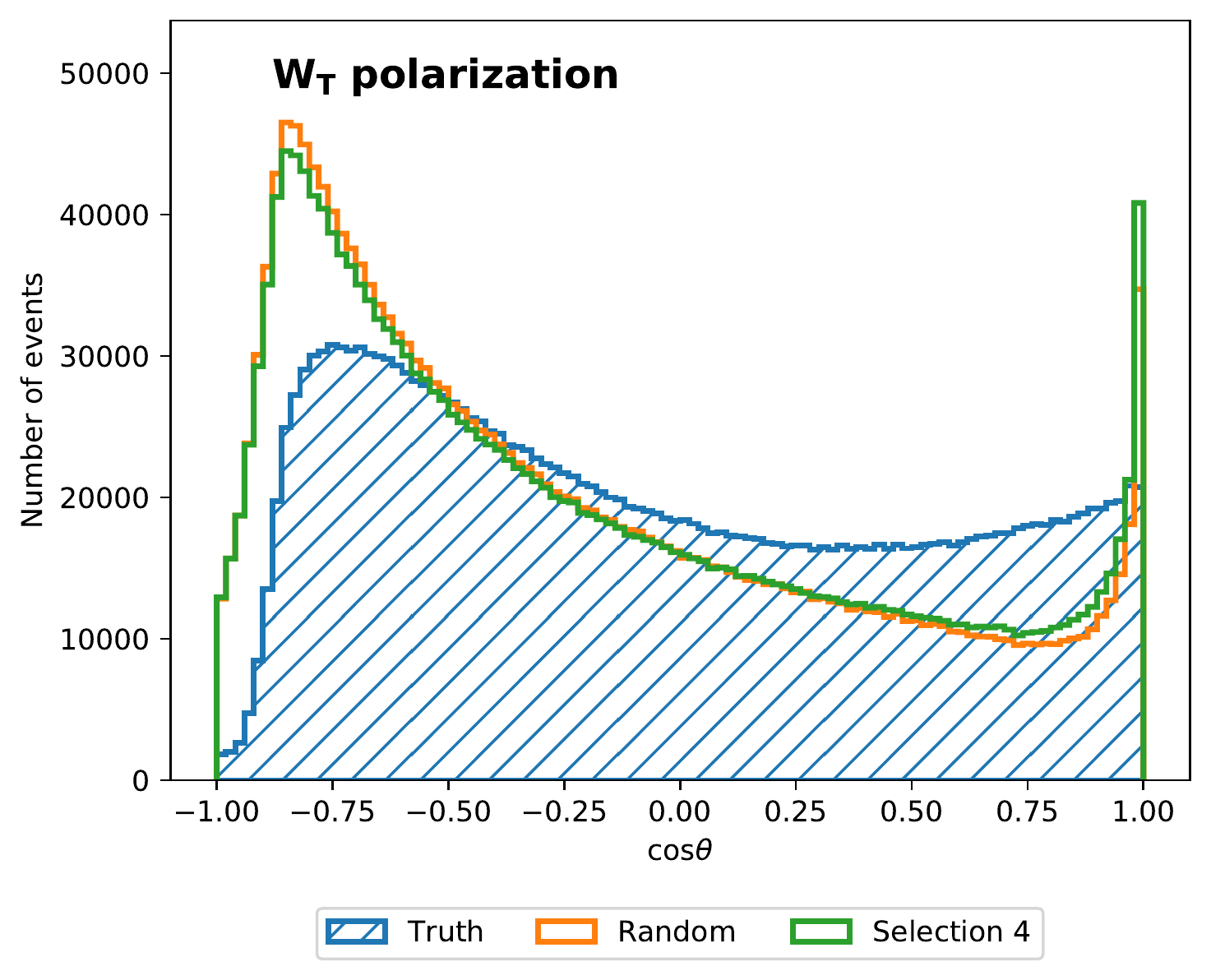}\\
\caption{MAOS reconstructed inclusive lepton (electron and muon) angular distribution in the $\mathrm{W}$ boson rest frame, compared to true parton level distribution, in the fully-leptonic channel. 
	The top plot shows the purely longitudinal ($\mathrm{W}_{\text{L}}$) and the bottom plot the purely transverse ($\mathrm{W}_{\text{T}}$) scenario. 
	Longitudinal neutrino momenta are calculated from the $\mathrm{W}$ mass constraint, using random solution and selection 4 strategies, described in Sect. ~\ref{subsec:semil}. Truth distributions use the generator-level (true) $\cos\vartheta$.}
\label{fig:cosmaos}
\end{figure}

In general, there are two classes of the solutions to the minimization problem (\ref{Eq:MAOS}) - \emph{balanced} and \emph{unbalanced}.
The difference between the two cases is diagrammatically shown in Fig.~\ref{fig:maossolution}. 
When the masses of the final state particles are negligible with respect to the $\mathrm{W}$ boson mass, the unbalanced case is not feasible and the minimum has to lie on the intersection of the $m^{(1)}_{\text{T}}(\vec{p}_1)$ and $m^{(2)}_{\text{T}}(\vec{p}_{\text{T}} - \vec{p}_1)$. 
In case of the leptonic $\mathrm{W}$ boson decay we find ourselves in this regime.
The intersection curve can be expressed analytically, while the $m_{\text{T}2}$ solution has to be estimated numerically along this curve. 
The derivation of an analytic solution for the intersection curve can be found in \ref{sec:app_maos}. 

After the transverse momenta of the two neutrinos have been determined, the subsequent kinematic reconstruction steps of the longitudinal neutrino momenta follow the same approach as in the semi-leptonic case described in Section~\ref{subsec:semil} for each $\mathrm{W}$ boson separately.
The performance is evaluated on purely longitudinally ($\mathrm{W}_{\text{L}}$) and purely transversely  ($\mathrm{W}_{\text{T}}$) polarized samples with both $\mathrm{W}$ bosons decaying leptonically, and therefore neutrino distributions can be shown inclusively for both $\mathrm{W}$ bosons in the event.
In Fig.~\ref{fig:pmaos} the success in the reconstruction of the neutrino momenta using the MAOS approach can be observed at the truth (generator) level. Ratios between the reconstructed  longitudinal and transverse neutrino momentum components $p_{\nu \text{L,T}}$ and true (generator-level) values $p_{\nu \text{L,T}}^\mathrm{truth}$ are shown inclusively for both neutrinos, since a complete symmetry is expected. 
A comparison of the true and reconstructed lepton angular distributions in the $\mathrm{W}$ boson rest frame for the longitudinally and transversely polarized samples is shown in Fig.~\ref{fig:cosmaos}.

To sum up: in the fully-leptonic case, the kinematic reconstruction of neutrino momenta is more complicated.
While the MAOS framework we used performs relatively well compared to other (non-machine-learning) approaches, it turns out that $\mathrm{W}$ polarization separation in the lepton angular distributions (evaluated in the reconstructed  $\mathrm{W}$ reference frame) is still quite poor, obscuring the differences between transverse and longitudinal polarizations.  
Consequently, as in the semi-leptonic case, trying a machine learning approach again seems worthy of investigation.

\section{Extracting polarization fractions using Neural Network}
\label{sec:dnn}

As already stated, one of the goals of this work is the investigation of a deep-learning multivariate technique to reconstruct the $\mathrm{W}$ boson rest frame(s) and consequently the polarization fractions.
Particular attention has been dedicated to the determination of the optimal network architecture and configuration, in terms of performance and reliability. 

Abundant literature (see, for instance, \cite{DNN:hepdnn1} and \cite{DNN:hepdnn2}) exists about machine learning (ML) and deep learning (DL) applications in high energy physics studies, including the specific case of VBS addressed here (\cite{dnnzz:PhysRevD.100.116010} and \cite{PhysRevD.93.094033}) where the authors addressed the problem using a different approach.

In the present work, we exploit a DL technique, namely Deep Neural Networks (DNN) developed, using two different approaches: Binary classification 
(Section~\ref{subsec:NNBin}) and Regression (Section~\ref{subsec:NNReg}). 
These two  approaches, explained in the following sections, make use of the same dataset for train, test and validation phase, the only difference between the two being the neural network problem formulation.
All details about dataset dimension, variables and assumptions can be found in \ref{sec:nndet}.

\begin{figure}
	\includegraphics[scale=0.55]{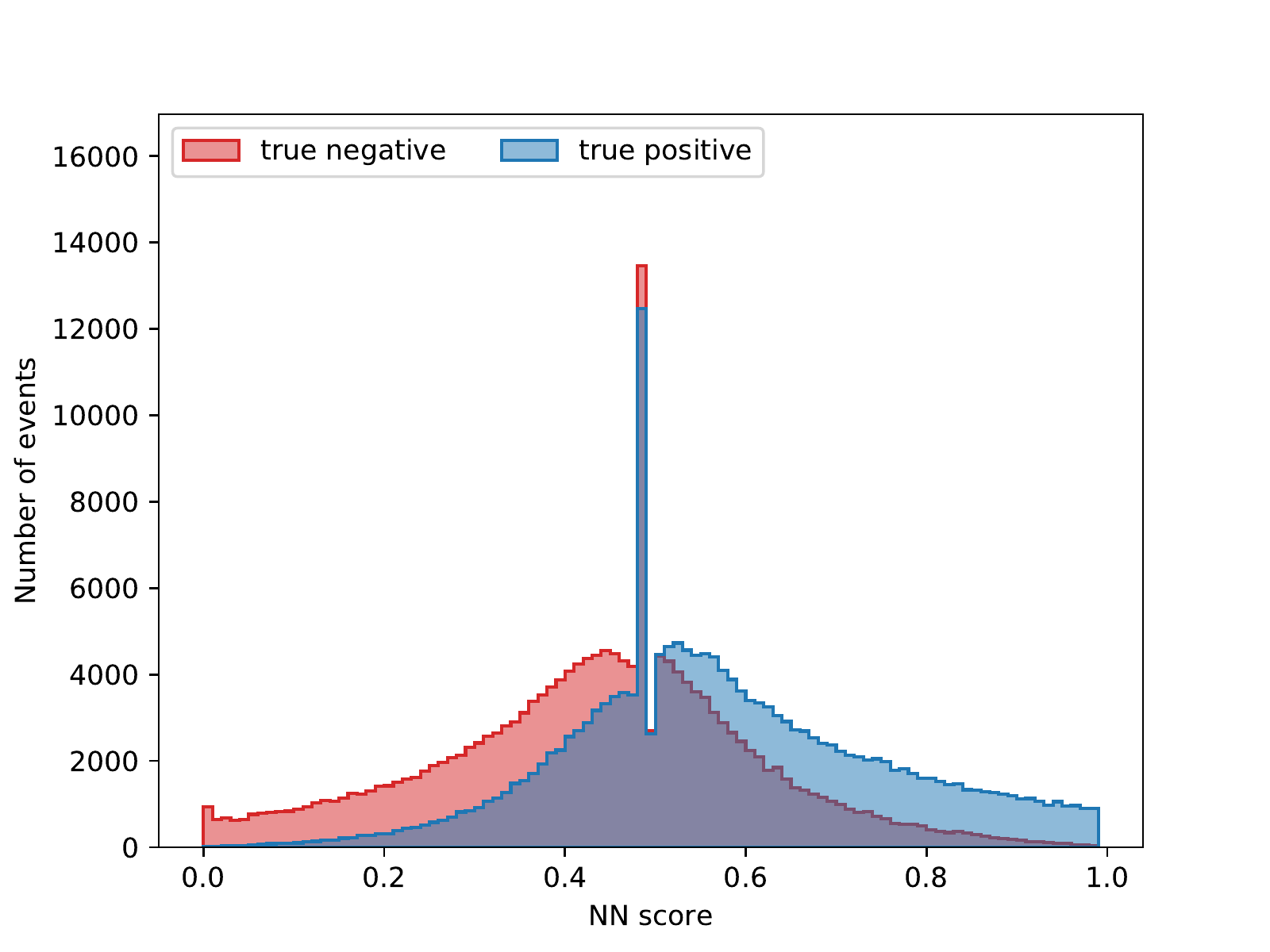}
	\caption{The DNN binary classification score in the semi-leptonic channel, determining the appropriate sign to use in neutrino momentum reconstruction is shown separately for events with true positive and true negative solutions, evaluated on events with transverse $\mathrm{W}$ polarization.
	The  DNN model is composed of 4 hidden layers and 60 neurons in each of them, the training was performed on an unpolarized \texttt{PHANTOM} sample.
	The sharp peak at 0.5 represents events for which the classifier cannot make a decision.}
	\label{fig:binscore} 
\end{figure}
\begin{figure}
	\includegraphics[scale=0.55]{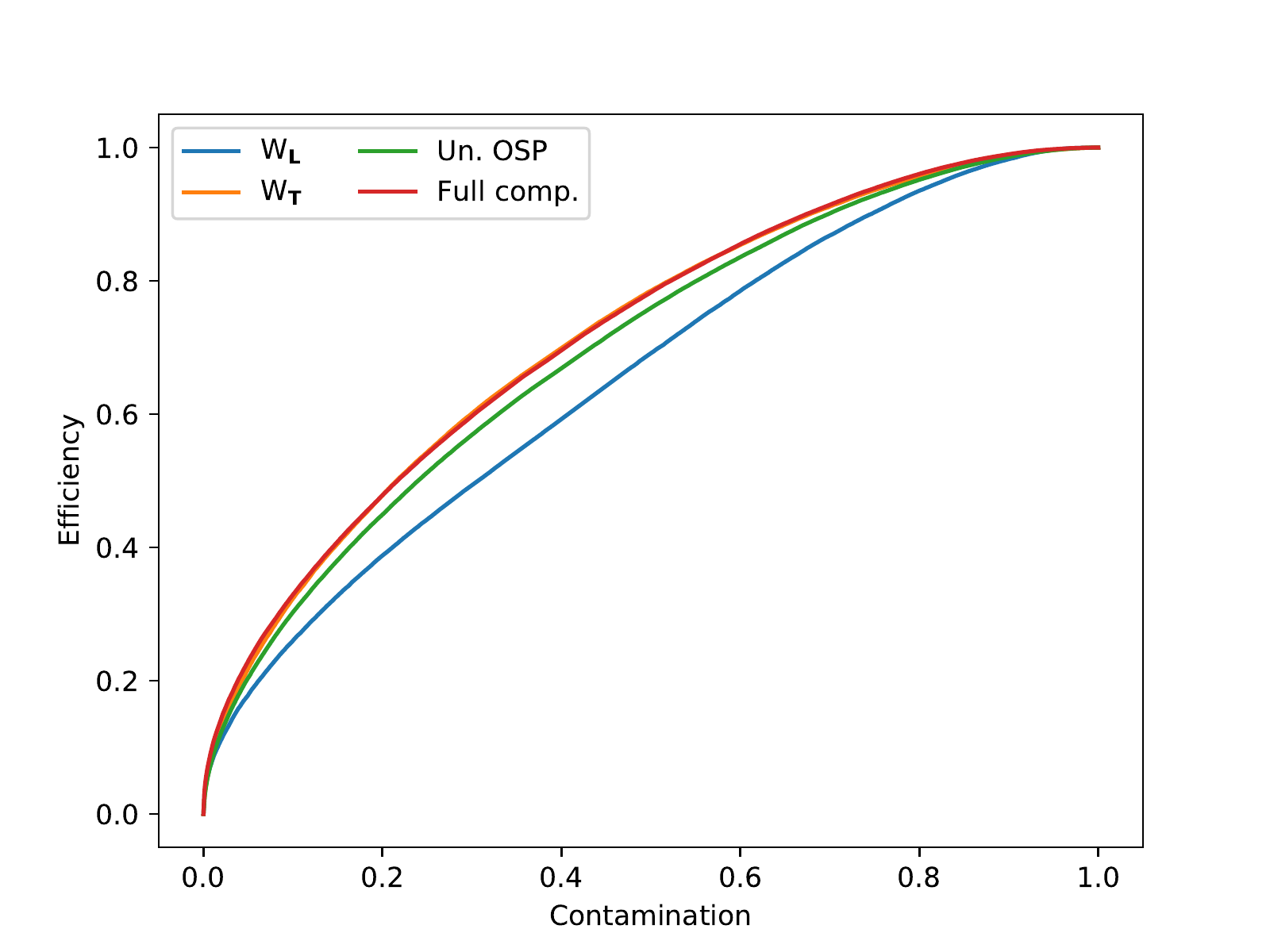}
	\caption{ROC curves in the semi-leptonic channel, representing the fraction of correct positive solution choices (efficiency) versus the fraction of negative solutions falsely labeled as positive (contamination), obtained on \texttt{PHANTOM} samples containing a purely longitudinally polarized ($\mathrm{W}_{\text{L}}$) leptonically decaying $\mathrm{W}$ boson, a purely transversely polarized ($\mathrm{W}_{\text{T}}$) boson, an unpolarized $\mathrm{W}$ boson generated in the OSP {\tt PHANTOM} framework (Un. OSP) and the full computation with an unpolarized $\mathrm{W}$ boson and with non-resonant production modes included (Full comp.) are shown.  The optimal working point is defined as the threshold giving the shortest distance to point (0,1). The $\mathrm{W}_{\text{T}}$ result coincides almost fully with the Full comp. and is consequently not visible in the plot.}
	\label{fig:roccurve}
\end{figure}

\subsection{Binary classification technique}
\label{subsec:NNBin}

The first DNN implementation adopts a binary classifier to determine the correct solution of Eq.~\ref{eq:soeq}, and then uses it to reconstruct the $\mathrm{W}$ boson reference frame. Binary classifier is a neural network with one neuron in the output layer. The code deduces which of the two ($+/-$) solutions is chosen by nature (addressed in the following as the \emph{correct solution}), based on training variables. In this way the kinematics of the boson decay is fully reconstructed, as in Section~\ref{subsec:semil} but with a set of selection criteria configured by the DL approach.

The DNN input is a set of 27 variables, and includes simple variables related to the momentum and geometrical disposition of the physical objects as well as more complex objects that combine several features. The detailed list is given in \ref{sec:nndet}.

Correct solutions are determined event by event by analyzing \texttt{PHANTOM} events. In our DNN, a label is assigned on event level basis following this criterion: events with negative (positive) sign giving the result closer to truth $p_{\nu \text{L}}$ are assigned label 0 (1), events with negative discriminant are discarded.\footnotemark
 
 \footnotetext{In this approximation the mass of $\mathrm{W}$ boson is fixed at the value of $80.38$ GeV, according to the Particle Data Book \cite{PDG:PhysRevD.98.030001}.}
 
 Our DNN is trained on truth solution labels to return score
 bounded between 0 and 1. The score distribution, obtained when training on the transversely polarized \texttt{PHANTOM} sample is shown in Fig.~\ref{fig:binscore}. Optimal working points for classification have been selected for each model based on the Receiver Operating Characteristic (ROC) curves \cite{DNN:roc}. An example of ROC curves for the DNN model using 60 neurons and 
 4 hidden layers is shown in Fig.~\ref{fig:roccurve}. Values of area under curve (AUC) and fractions of correct solution choices are summarized in 
 Table~\ref{tab:aucfrac}. 

\begin{center}
\begin{table}
	\caption{Classifier performance in the semi-leptonic channel, using 60 neurons and 4 hidden layers. Results obtained on samples containing a purely longitudinally polarized ($\mathrm{W}_{\text{L}}$) leptonically decaying $\mathrm{W}$ boson, a purely transversely polarized ($\mathrm{W}_{\text{T}}$) boson, an unpolarized $\mathrm{W}$ boson generated in the OSP {\tt PHANTOM} framework (Un. OSP) and the full computation with an unpolarized $\mathrm{W}$ boson and with non-resonant production modes included (Full comp.) are shown.
	}
	\label{tab:aucfrac} 

	\begin{tabular}{lllll}
		\hline\noalign{\smallskip}
		 & $\mathrm{W}_{\text{L}}$ & $\mathrm{W}_{\text{T}}$ & Un. OSP & Full comp. \\
		\noalign{\smallskip}\hline\noalign{\smallskip}
		Working point & 0.497 & 0.498 & 0.500 & 0.491 \\
		AUC & 0.655  & 0.716 & 0.698 & 0.717 \\
		Corr. choices & 0.596 & 0.651 & 0.636 & 0.650 \\
		\noalign{\smallskip}\hline
	\end{tabular}
\end{table}
\end{center}

\begin{figure}
	\includegraphics[scale=0.55]{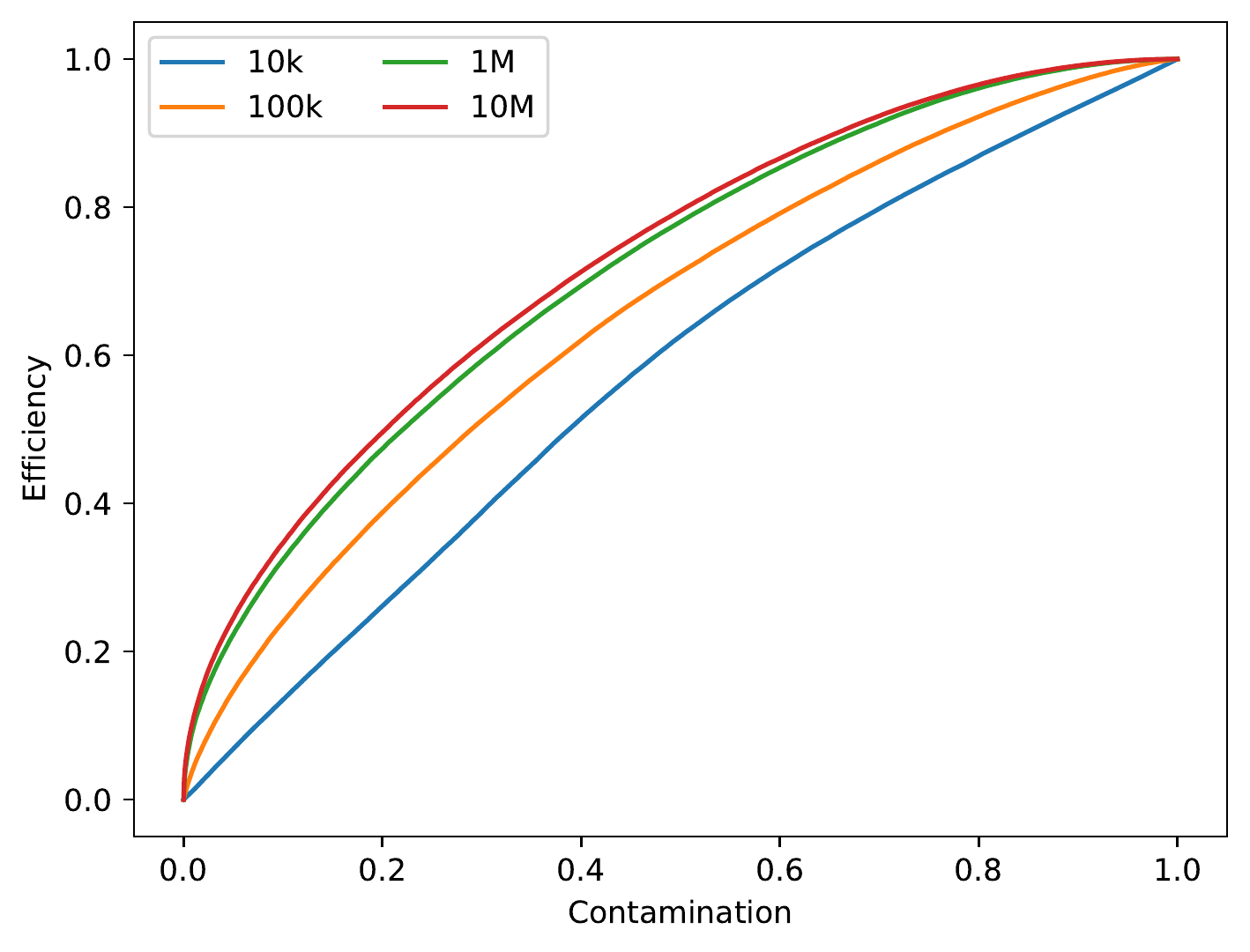}
	\caption{ROC curve in the semi-leptonic channel, showing dependency on the training unpolarized OSP dataset size ($10^4$, $10^5$, $10^6$, $10^7$ events). Number of neurons is fixed to 60 and number of hidden layers to 4.}
	\label{fig:roccompare} 
\end{figure}

A dedicated study has been made to evaluate the dependence of the DNN on the dimension of the training set. 
Different sizes of the training sets have been considered, spanning from 10k events to 10M events. The resulting ROC curves are shown in 
Fig.~\ref{fig:roccompare}. 
One can see that the performance improves with the increase of the training set size, but seems to converge quite well at size 1M, since the difference between 1M and 10M can be considered negligible. 
Given this feature, we chose 1M training set size, to perform further optimizations of the DNN parameters.

\begin{figure}
 \includegraphics[width=0.45\textwidth]{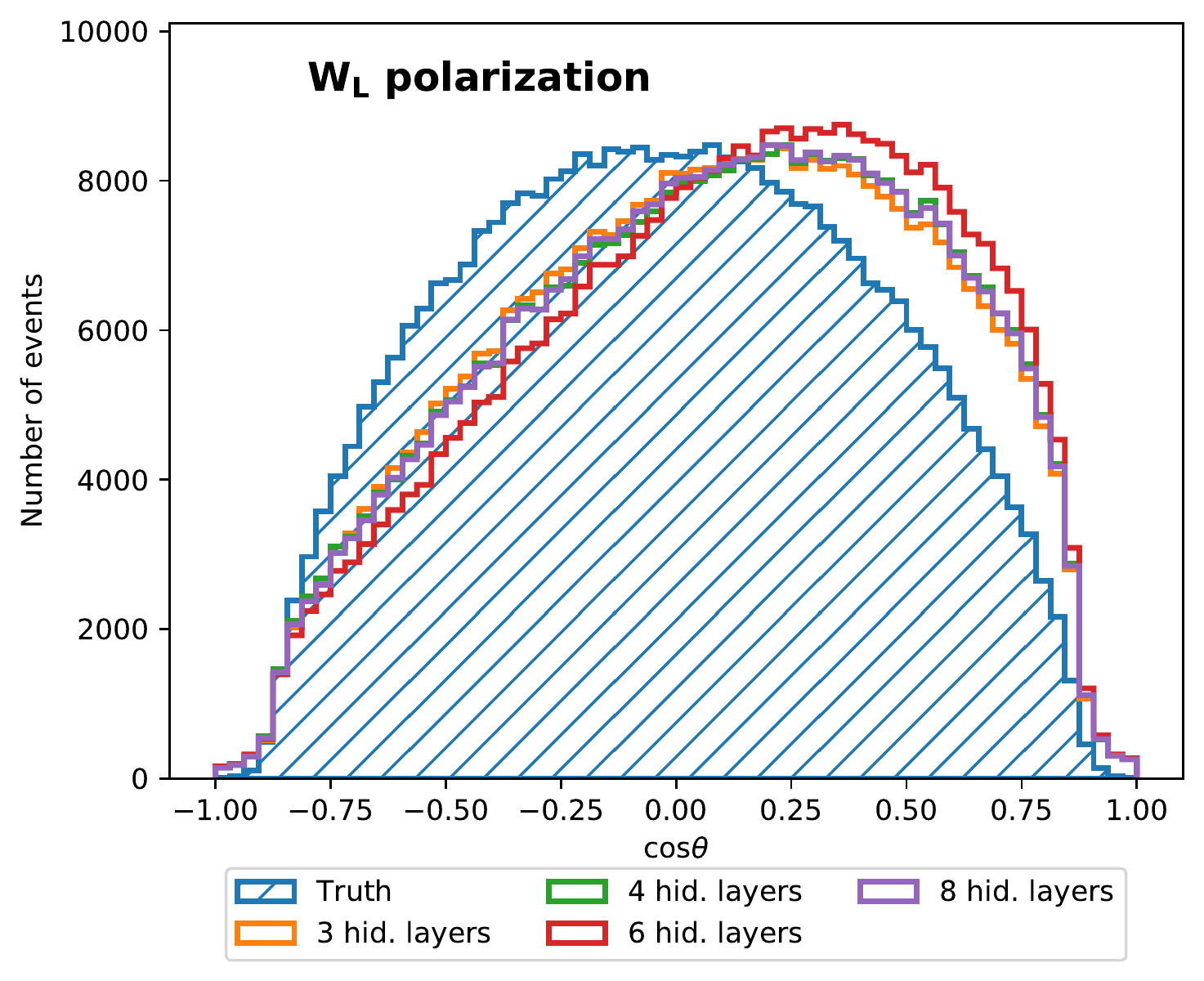}
 \includegraphics[width=0.45\textwidth]{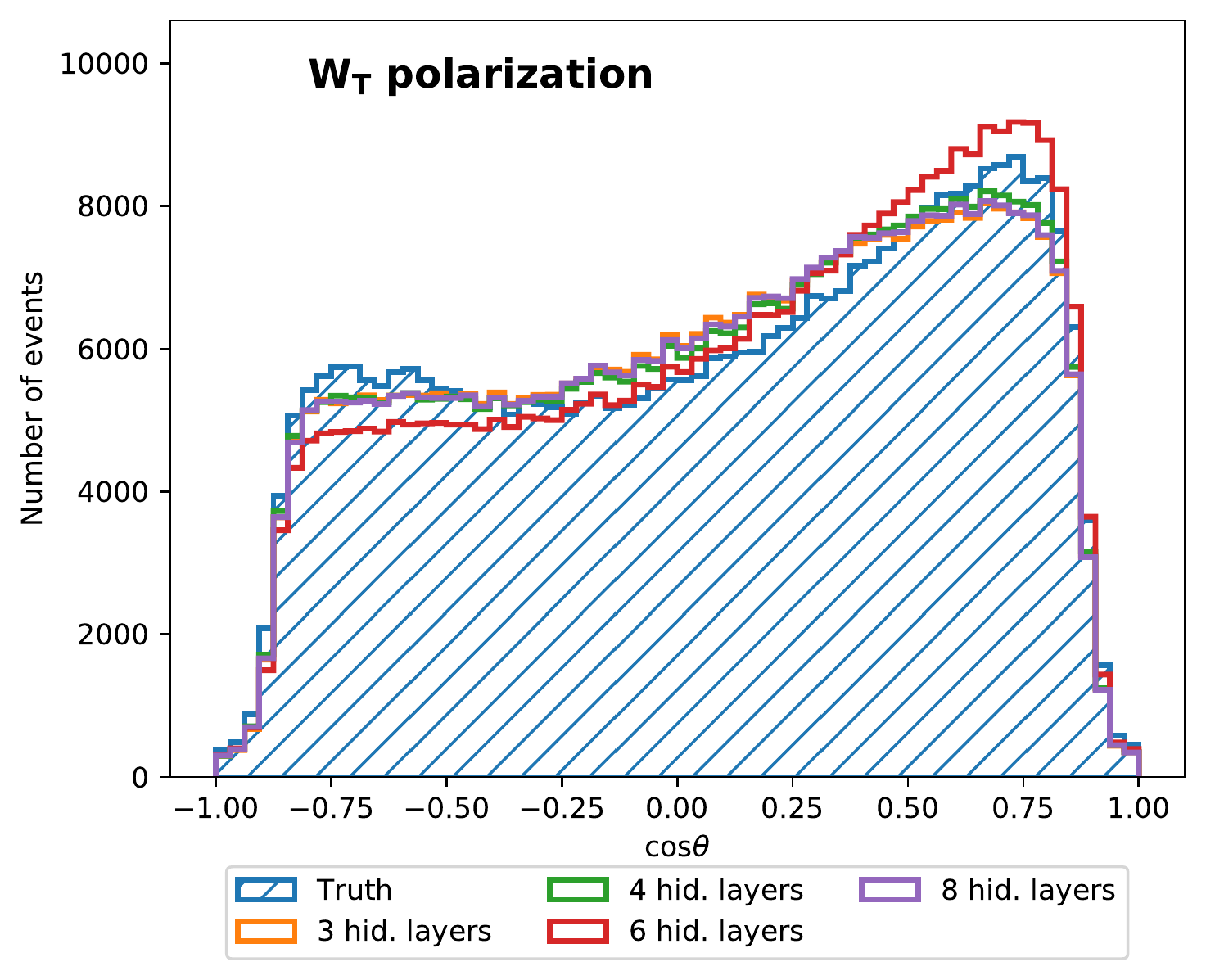}\\
\caption{Angular distribution in the $\mathrm{W}$ boson rest frame as a function of $\cos\vartheta$, obtained using the binary classification technique, in the semi-leptonic channel.
Parton-level (truth) distributions are shown for comparison, using the generator-level (true) $\cos\vartheta$.
The top plot shows results obtained assuming the purely longitudinal ($\mathrm{W}_{\text{L}}$) and the bottom plot the purely transverse ($\mathrm{W}_{\text{T}}$) scenario.
	The results for a DNN with a fixed number of neurons (60) and different hidden-layer configurations are shown.}
\label{fig:bindist}       
\end{figure}

Following Eq.~\ref{eq:soeq}, $p_{\nu \text{L}}$ and $E_{\nu}$ in the laboratory frame are calculated. Finally the lepton 
momentum is boosted into reconstructed $\mathrm{W}$ rest frame, defined by the vector sum of neutrino and lepton momentum. The resulting angular distribution in the $\mathrm{W}$ boson rest frame, obtained using the Binary classification technique, is shown in Fig.~\ref{fig:bindist} as a function of $\cos\vartheta$, for a DNN with a fixed number of neurons (60) and different hidden-layer configurations.


\begin{figure}
	\includegraphics[width=0.45\textwidth]{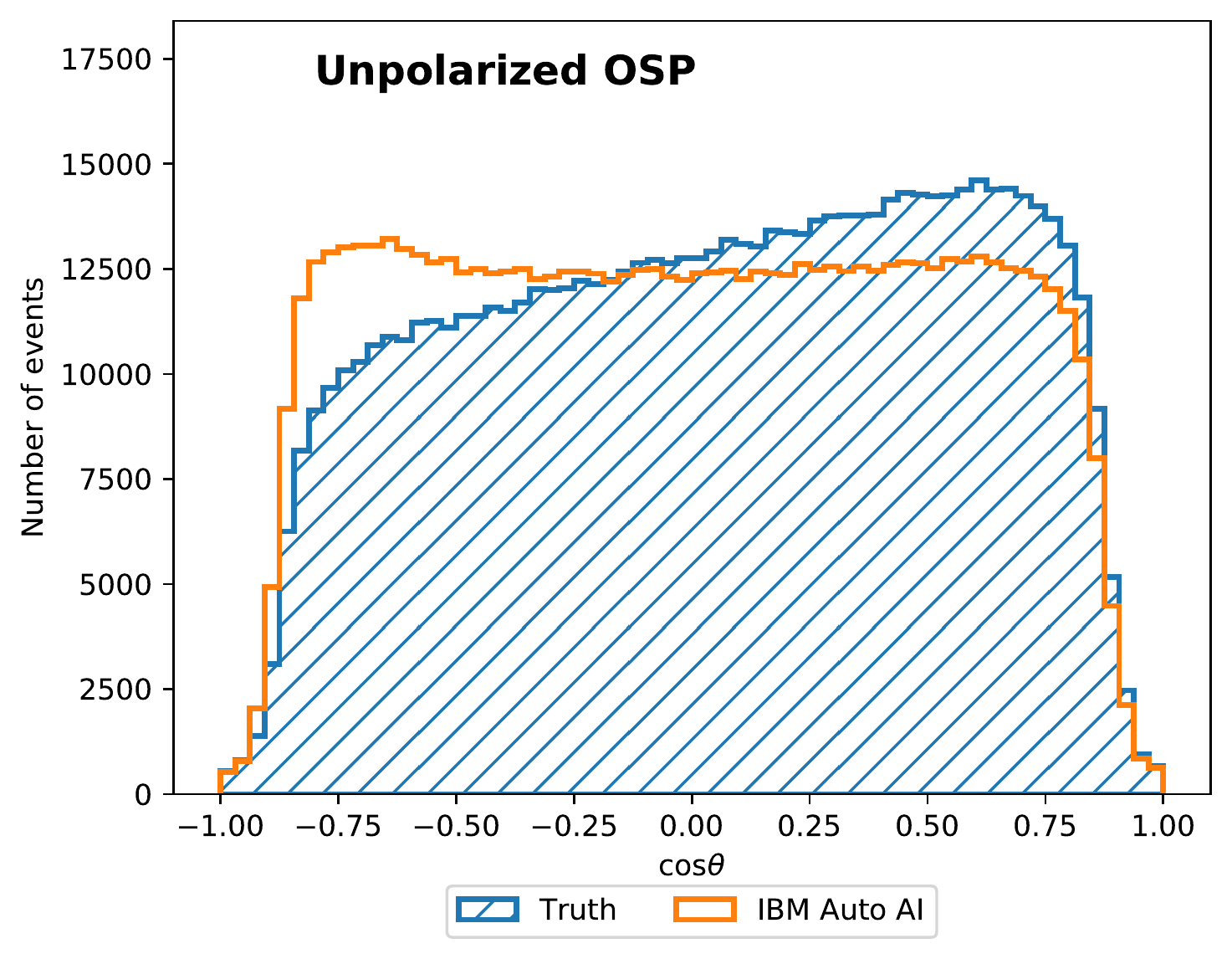}
	\caption{Angular distribution in the $\mathrm{W}$ boson rest frame as a function of $\cos\vartheta$, obtained using the IBM auto AI tool as a binary classificator. Parton-level (truth) distributions are shown for comparison, using the generator-level (true) $\cos\vartheta$.}
	\label{fig:autoaidist}       
\end{figure}

In addition to the standard open-source platforms we used a specific one available on IBM Cloud, named AutoAI, that exploits artificial intelligence routines to generate candidate model pipelines customized for your predictive modeling problem.
These model pipelines are created iteratively as AutoAI analyses datasets and discovers data transformations, algorithms, and parameter settings that work best for the problem setting. It covers all the steps of a typical
machine learning project from data pre-processing, to model selection, feature optimization and model deployment.
In particular it automatically detects and categorizes features based on data type, such as categorical or numerical. Then it transforms raw data into the combination
 of features that best represents the problem to achieve the most accurate prediction. AutoAI uses a unique approach that explores various feature construction 
 choices in a structured, non-exhaustive manner, while progressively maximizing model accuracy using reinforcement learning.
 Finally, a hyper-pa\-ra\-me\-ter optimization step refines the best performing model pipelines. AutoAI uses a novel hyper-parameter optimization algorithm optimized for costly function evaluations such as model training and scoring that are typical in machine learning.
  This approach enables fast convergence to a good solution despite long evaluation times of each iteration.
Comparing Figures \ref{fig:bindist} and \ref{fig:autoaidist}, we see that there is no further improvements in our binary classification as we extracted all possible information belonging to training data.
We made this comparison for binary classification, for which the AUC value represent a clear statistical indicator, although we got similar conclusion also for the technique that will be discussed in the next paragraph.

\subsection{Regression technique}
\label{subsec:NNReg}

A second approach can be applied. Instead of using the neural network to address the solution of Eq.~\ref{eq:soeq}, we could train the algorithm to determine directly the reconstructed angular distributions of the lepton in $\mathrm{W}$ rest frame,
  with respect to the $\mathrm{W}$ direction in the laboratory frame.
The DNN input is again a set of 27 variables, from simple variables related to the momentum and geometrical disposition of the physical objects
to more complex objects that combine several features. The detailed list is given in \ref{sec:nndet}. Differently from the binary classification approach, here the DNN is trained with the true value of the $\cos\vartheta$ variable, which it aims at reproducing. 
With this approach, we tested as many different neural network topologies as in the case of binary classification, albeit changing the relevant parameters that convert a classification problem into a regression.

\begin{figure}
  \includegraphics[width=0.45\textwidth]{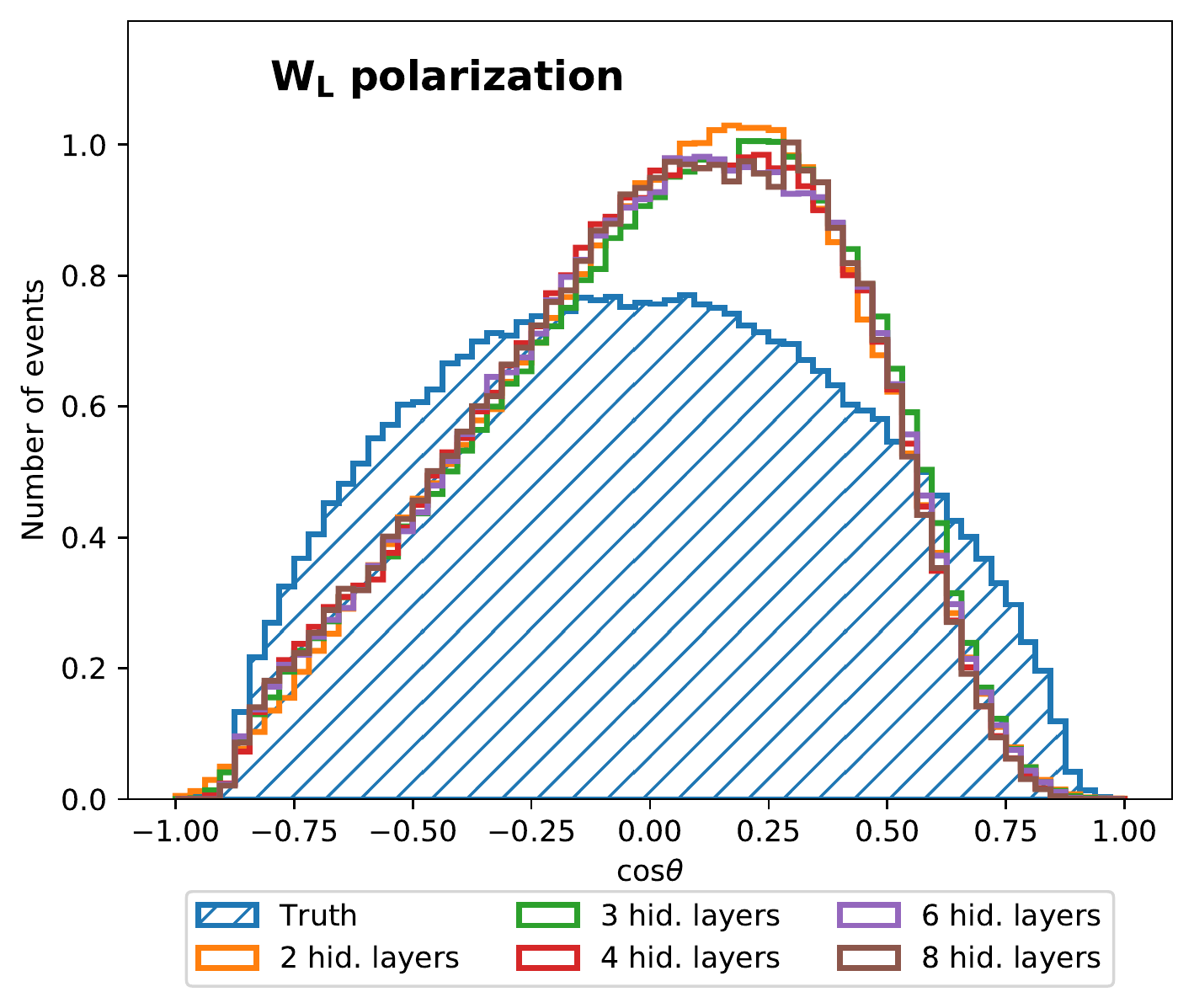}
  \includegraphics[width=0.45\textwidth]{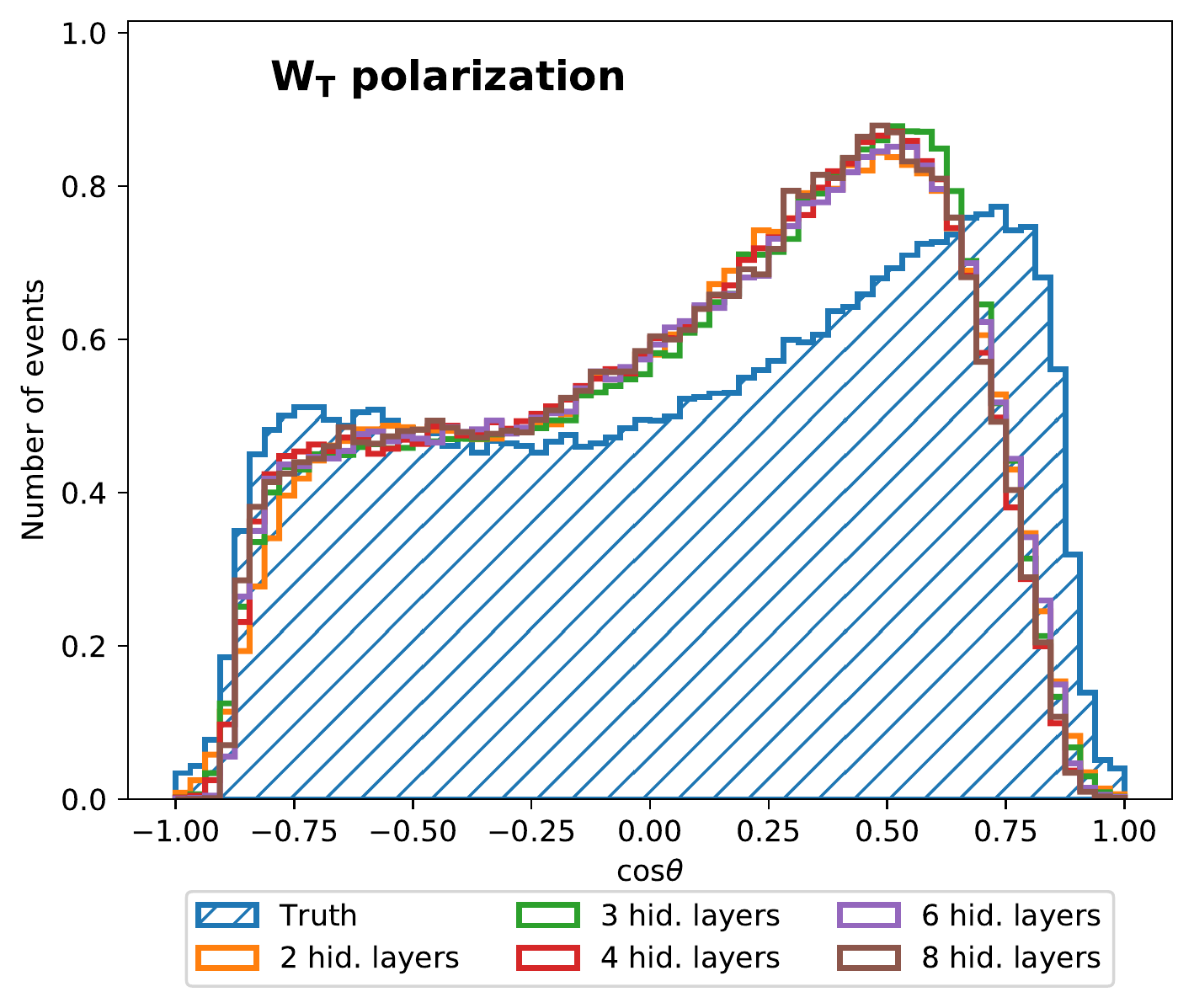}
\caption{Angular distribution in the $\mathrm{W}$ boson rest frame as a function of $\cos\vartheta$, obtained using $\cos\vartheta$ regression, in the semi-leptonic channel. Parton-level (truth) distributions are shown for comparison, using the generator-level (true) $\cos\vartheta$.
The top plot shows results obtained assuming the purely longitudinal ($\mathrm{W}_{\text{L}}$) and the bottom plot the purely transverse ($\mathrm{W}_{\text{T}}$) scenario. Results for a DNN with a fixed number of neurons (60) and different hidden-layer configurations are shown.}
\label{fig:regdist}       
\end{figure}

Given the information about the correct value of the $\cos\vartheta$ from the regression DNN approach, the resulting angular distribution in the $\mathrm{W}$ boson rest frame as a function of $\cos\vartheta$, is shown in Fig.~\ref{fig:regdist} for a DNN with a fixed number of neurons (60) and different hidden-layer configurations. To evaluate the performance of this approach, the distribution of the reconstructed $\cos\vartheta$ for longitudinally and transversely polarized samples angular distribution is compared to the truth values. 

\begin{figure}
\includegraphics[scale=0.55]{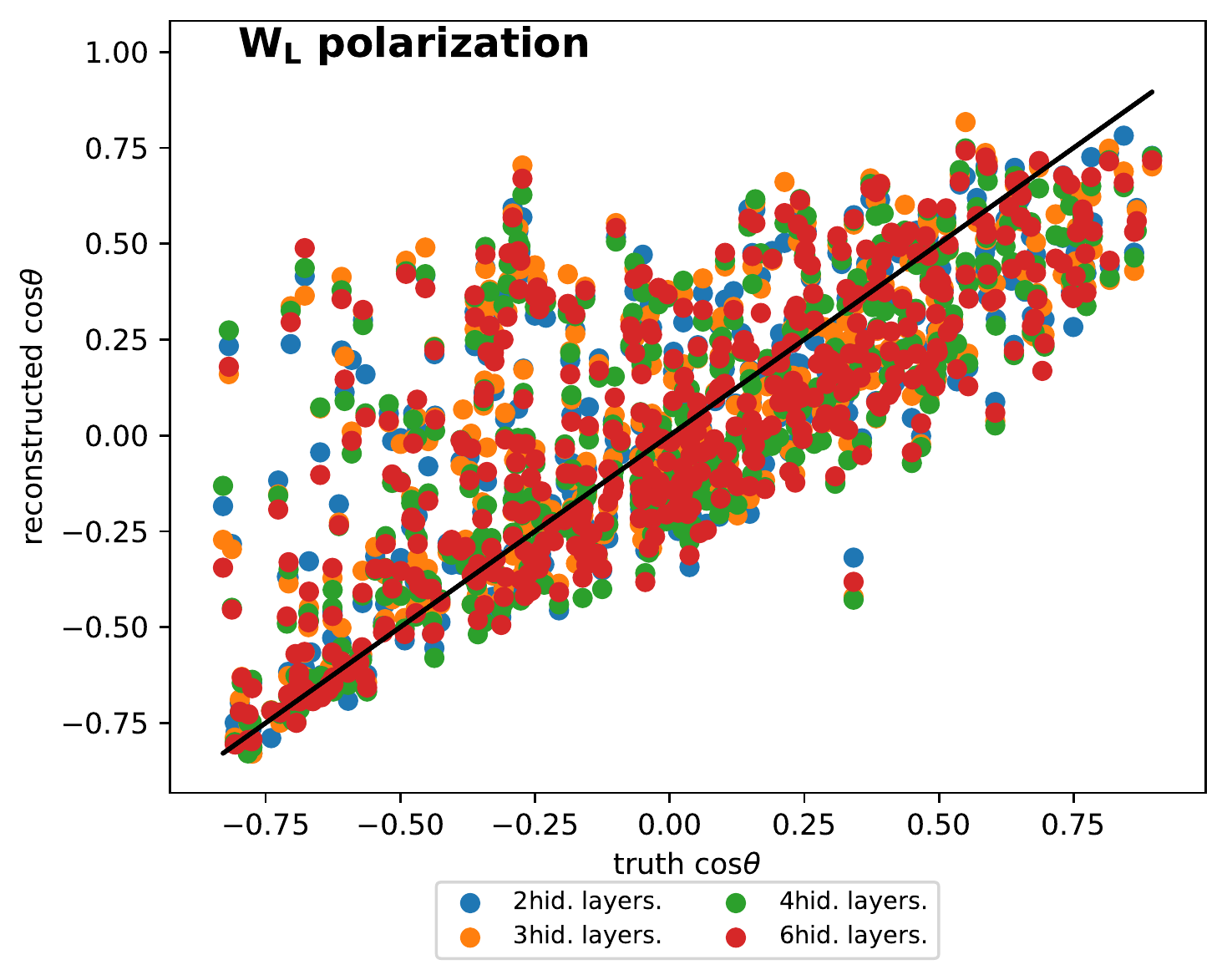}
\caption{Scatter plot: generator-level (true) $\cos\vartheta$ value w.r.t.\@ the reconstructed one in the semi-leptonic channel, using different network topologies.}
\label{fig:regscatt}    
\end{figure}

One additional way to evaluate the performance of the model for different neural network topologies is presented in Fig.~\ref{fig:regscatt},
 where the distance of points from the main diagonal represents the ability of the network to reconstruct $\cos\vartheta$.

\begin{figure}
\includegraphics[scale=0.55]{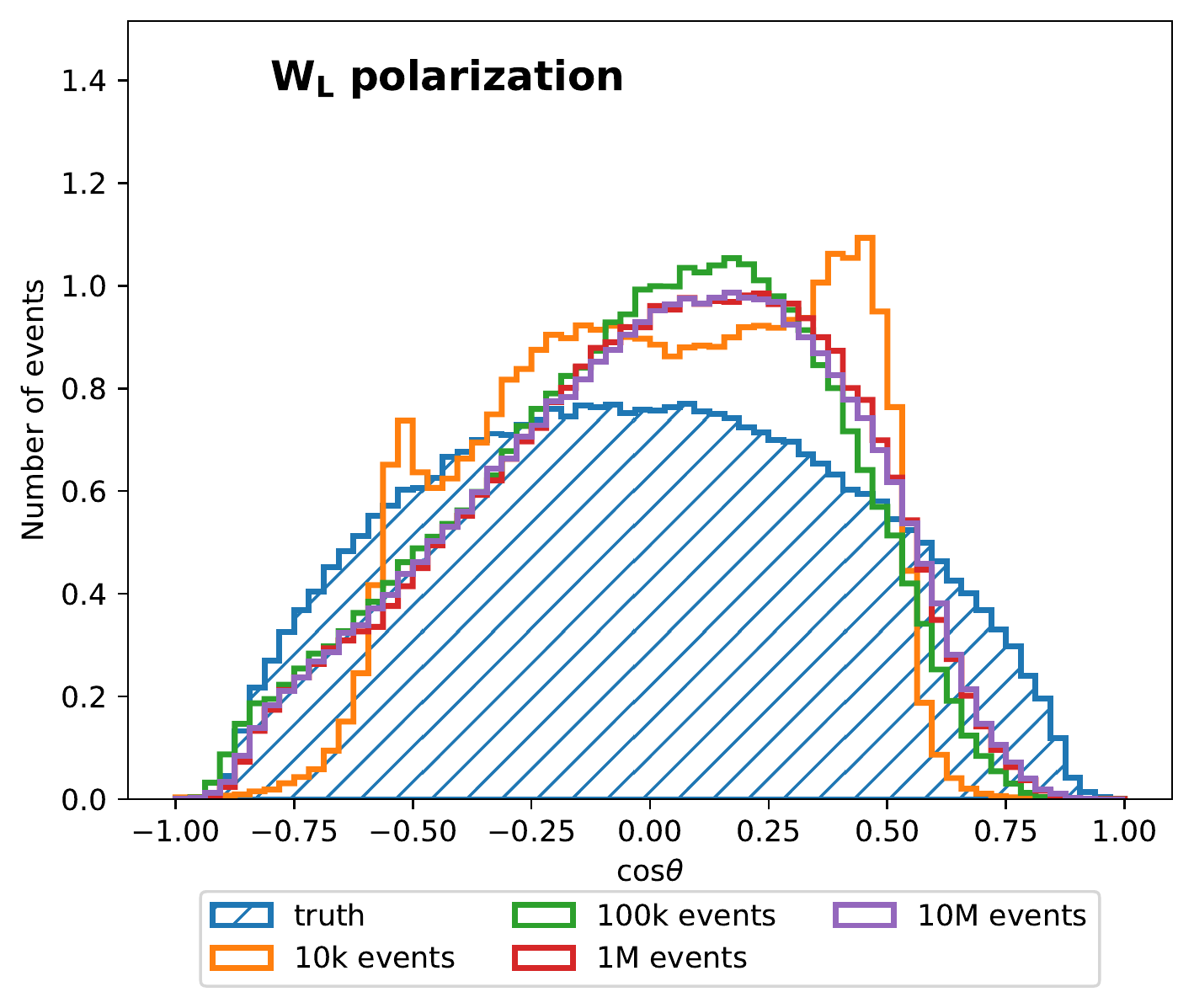}
\caption{Reconstructed lepton angular distribution in the semi-leptonic channel, for the longitudinal $\mathrm{W}$ boson polarization, as a function of $\cos\vartheta$ with a specific network topology (60 neurons, 4 hidden layers) for model trained with
different training dataset sizes ($10^4$, $10^5$, $10^6$, $10^7$ events). Parton-level (truth) distributions are shown for comparison, using the generator-level (true) $\cos\vartheta$. }
\label{fig:regcompare}       
\end{figure}

Similarly to binary classification, a dedicated study has been made to evaluate the dependence of the DNN on the dimension of the training set. 
Different sizes of the training sets have been considered, spanning from 500 events to 10M events.
Results are shown in Fig.~\ref{fig:regcompare}. As expected, the larger the training set size, the better the performance, although one should also consider the possibility of over-fitting for the DNN.

\subsection{Neural Network approach to Fully-leptonic Channel}
\label{sec:NN_Full}

Following the approach used for the semi-leptonic case, we applied the DNN study  to the fully-leptonic channel.
 From a technical perspective, this channel can be considered as a direct extension of the 
semi-leptonic case presented before, where only the regression approach can be considered.  
The fully-leptonic channel kinematics reconstruction can be supplemented with the MAOS algorithm, as described in Section~\ref{sec:fullep}. 
Since we want to check whether the information obtained from the MAOS could improve the DNN reconstruction, we included MAOS algorithm outputs among the training features, namely MAOS reconstructed neutrino transverse momenta ($p_{\text{T}}^{\nu_{\ell}}$ from Eq.~\ref{Eq:MAOS}) and $m_{\text{T2}}$. 
In the fully-leptonic channel, our methodology extends the possibility for the network to reconstruct  both
$\cos\vartheta$ for the muon and electron part directly and then combines these two contributions to get the total angular distribution of a certain polarization component. This approach will be referred to as a \emph{direct approach}, from here on.

\begin{figure}
	\includegraphics[width=0.45\textwidth]{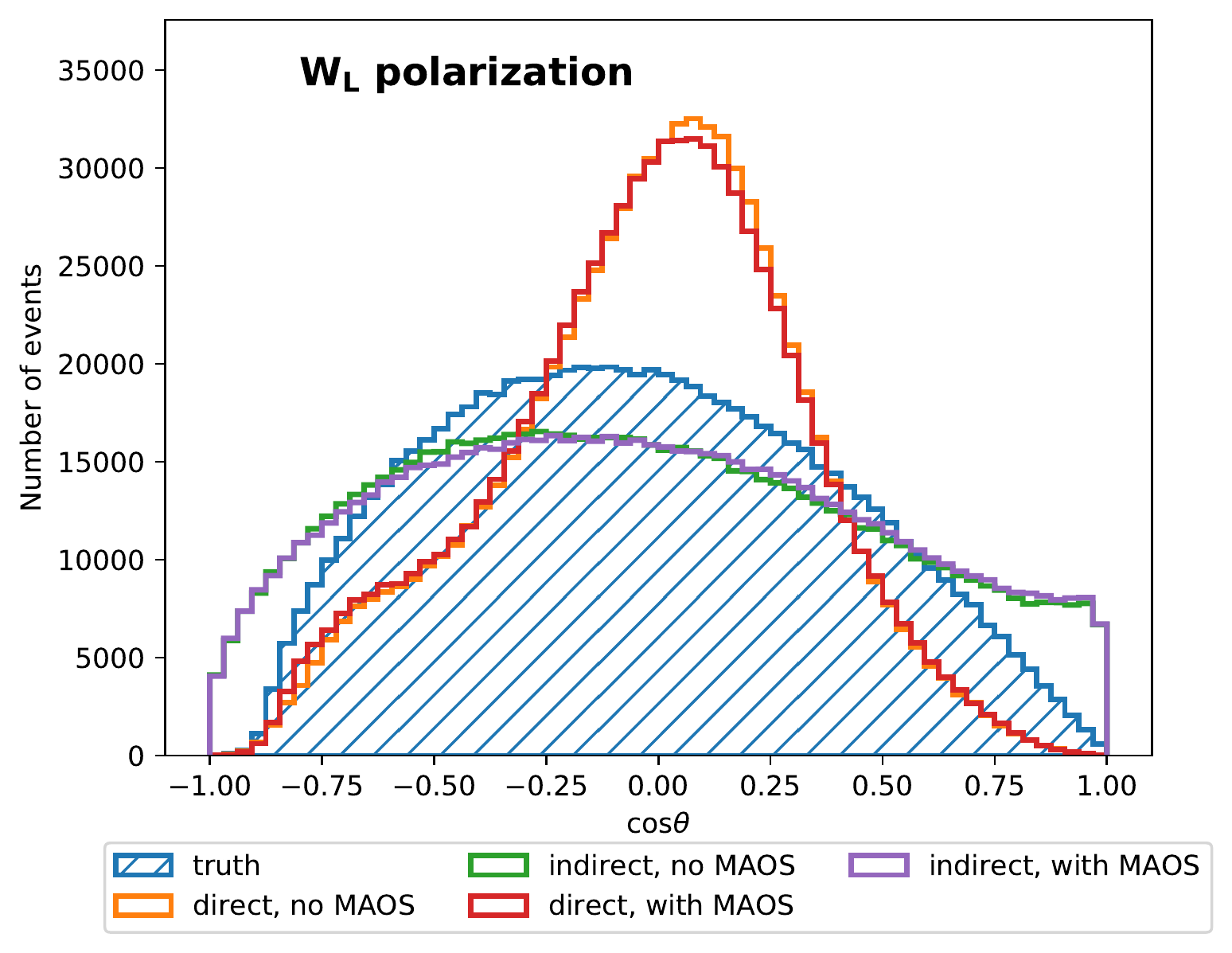}
	\includegraphics[width=0.45\textwidth]{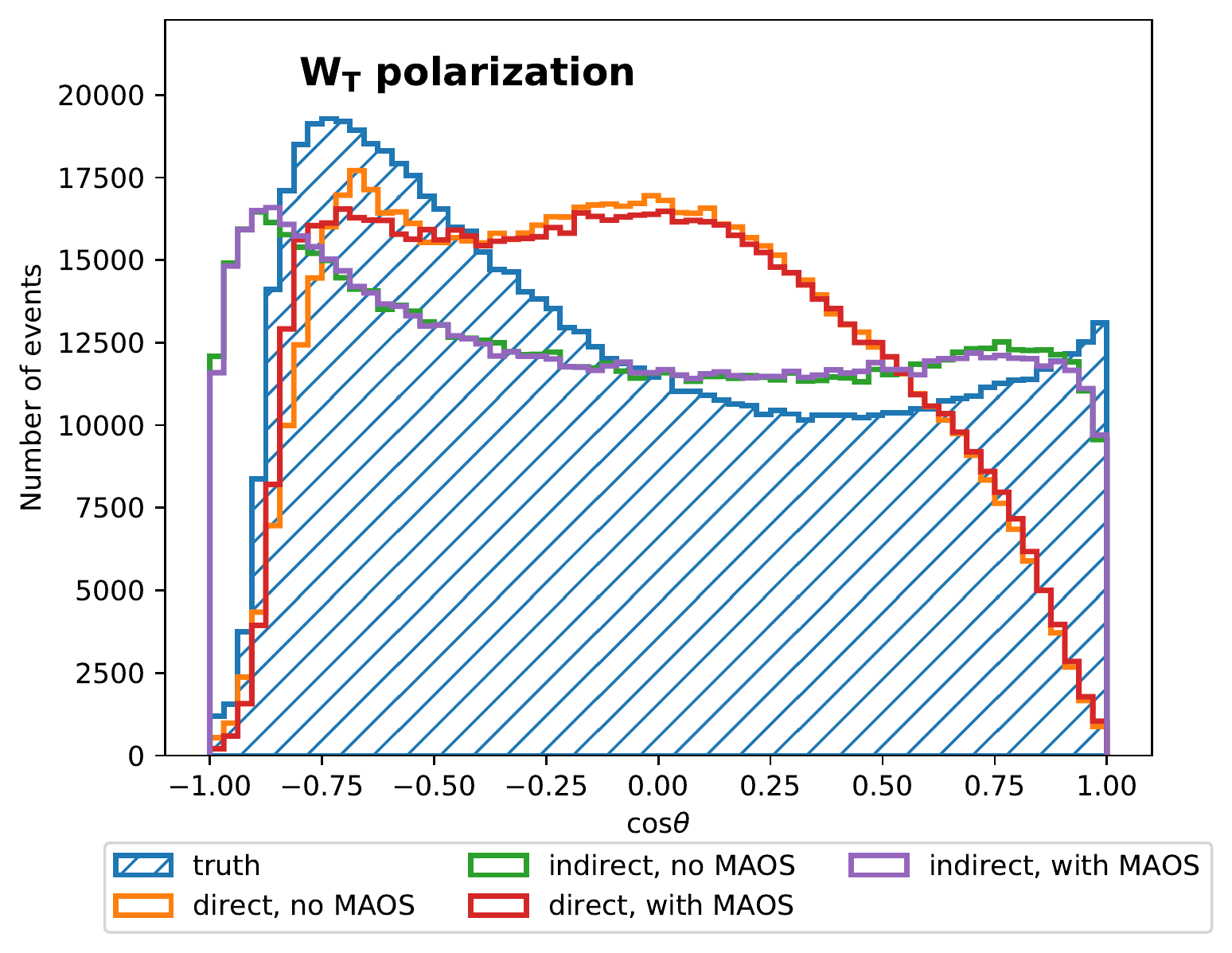}
	\caption{Shape of the reconstructed lepton (electron and muon) angular distribution in the $\mathrm{W}$ boson rest frame in the fully-leptonic channel. 
	Parton-level (truth) distributions are shown for comparison, using the generator-level (true) $\cos\vartheta$.
 The top plot shows the the (purely) longitudinally polarized $\mathrm{W}_{\text{L}}$  and the bottom plot the (purely) transversely polarized $\mathrm{W}_{\text{T}}$ sample. Performance of the direct and indirect approach can be appreciated qualitatively. 60 neurons and 3 hidden layers have been used for all the approaches. Other DNN topologies give similar results.}

	\label{fig:fullylepsummary}
\end{figure}

As another step in the process of improving our ability to extract a model for angular distribution we implemented also an \emph{indirect approach} by introducing an intermediate step: 
in this case the DNN training labels  are now the six components of neutrino momenta and the neural network has to solve a regression problem with a six-dimensional output. Once the DNN model outputs the reconstructed six neutrino components,
we calculate and reconstruct the $cos\vartheta$ distribution. A summary of the performance of both reconstruction approaches, with and without the inclusion of the MAOS quantities among the training features, is shown in Fig.~\ref{fig:fullylepsummary}. 

\begin{figure}
\includegraphics[scale=0.55]{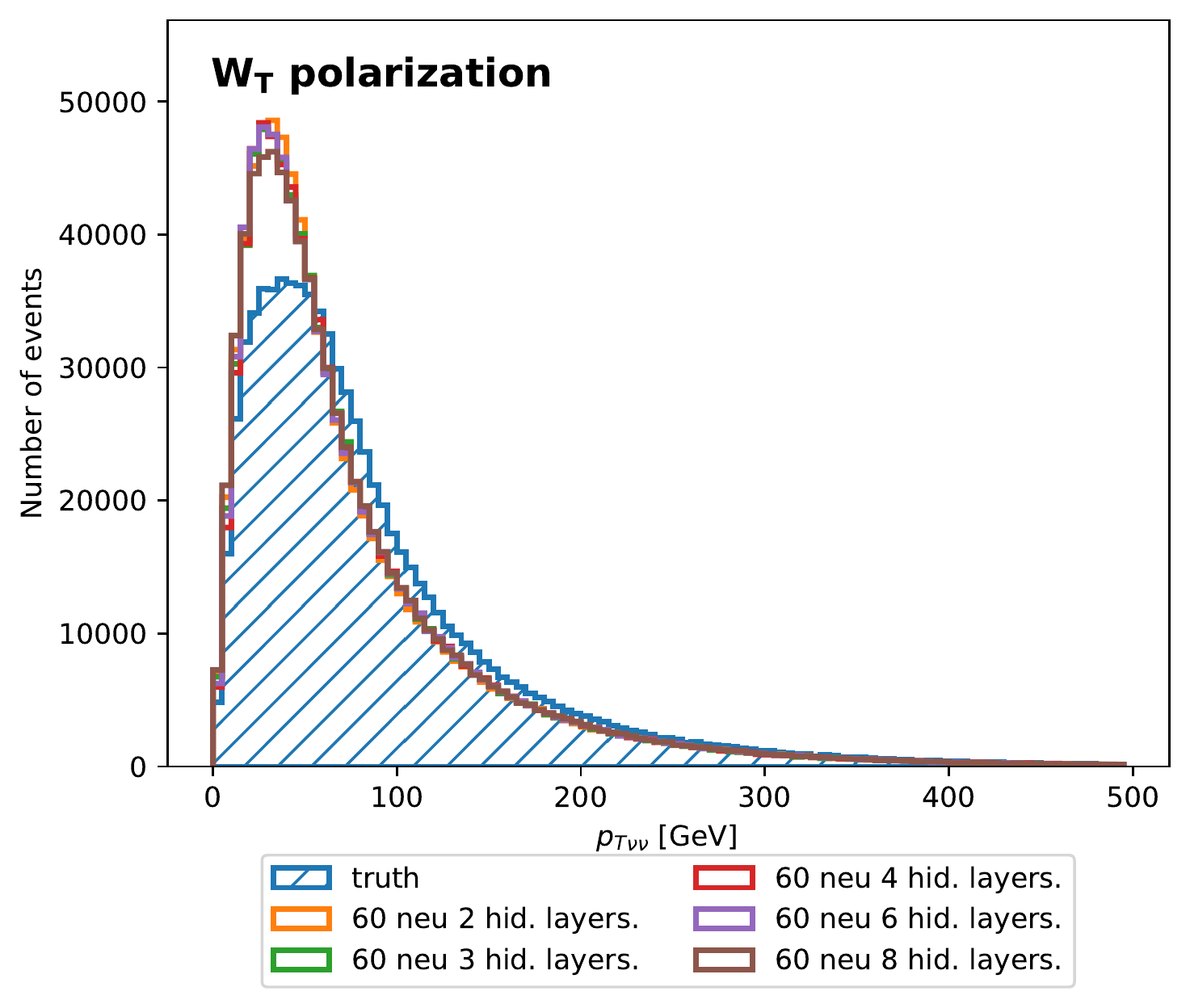}
\caption{Reconstruction of $p_{\text{T}\nu \nu}$ for different neural network topology with the same batch-size, applied on the fully-leptonic channel using a transversely polarized $\mathrm{W}_{\text{T}}$ scenario.  }
\label{fig:fulptvv}       
\end{figure}

As a further validation of the implemented reconstruction procedure, in Fig.~\ref{fig:fulptvv} we plot the reconstructed transverse momentum of the neutrino pair ($p_{\text{T}\nu \nu}$) that represents the total missing momentum of the process. From a deep learning perspective we tested more network topologies
with respect to the semi-leptonic case, as we noticed that some particular combinations of neurons and hidden layers were not able to learn. We realized that the loss functions in some cases were flat, or that we had to increase the number of training epochs
suggesting complex convergence structure. Evolution is in some cases represented by a negative exponential decay of the loss function and in other cases by a composition of step losses.

\subsection{Comparison and fit results}
\label{sec:Fit}

We estimated the goodness of a particular method in terms of its ability to extract the longitudinal polarization fraction from the unpolarized distribution. 
This is done by means of a maximum likelihood fit of the polarized angular distributions to the full computation.
The full computation does not only contain admixtures of the longitudinal and transverse polarization, but also interference among them and non-resonant contributions.
Since the interference can be constructive as well as destructive, we adopted an approach in which all the non-purely longitudinal components (trans\-ver\-se po\-la\-ri\-za\-tion, in\-ter\-fe\-ren\-ce and non\-re\-so\-nant con\-tri\-bu\-tion) are summed up and treated as a single distribution.
 They are obtained by subtraction of the longitudinal distribution from the full computation.
With this approach we avoid fitting histograms with negative entries that cannot be handled in the maximum likelihood fit.
From here on, the mixture of these three components will be referred to as the \emph{background} distribution.

In the fit, the full computation is hypothesized to be well described by the sum of the longitudinal and background distribution and its normalizations are chosen to be the fit parameters.
There is of course a very high anti-correlation between the fitted parameters because the benchmark is a direct sum of both.
The latter has as a consequence also the exact convergence of the best fit values to the truth polarization fractions, as predicted by the generator.

Interesting values are the confidence intervals (CI) for longitudinal and background normalization. 
The resulting normalization CI widths at 95\% CL using different sources of angular distributions (random solution choice, binary classification, regression and truth) are summarized in  Table~\ref{tab:fitCI}.\\
\begin{center}
\begin{table}
\caption{Summary of the longitudinal polarization fraction extraction from the fit of the background and longitudinal ($\mathrm{W}_{\text{L}}$)  distributions to the full computation. The background is defined as the difference between the full computation with an unpolarized $\mathrm{W}$ boson and with non-resonant production modes included (Full comp.) and the longitudinal ($\mathrm{W}_{\text{L}}$) contribution.
The best fit value, for all the reconstruction methods, matches the truth polarization fraction, $\sigma_{\text{L}}/\sigma_{\text{full}} = 0.257$, within the sub-percent level. CI widths at the 95 \% CL for background and longitudinal polarization fraction are reported. The numbers are obtained at 139.0 fb$^{-1}$ luminosity, which is the value of the integrated ATLAS luminosity for the 2015--2018 data taking period.}
\label{tab:fitCI}       
\begin{tabular}{llll}
\hline\noalign{\smallskip}
Type & Background  &  $\mathrm{W}_{\text{L}}$  \\
\noalign{\smallskip}\hline\noalign{\smallskip}
Random & 0.122 & 0.119 \\
Selection 1 & 0.137 & 0.135 \\
Selection 2 & 0.125 & 0.122 \\
Selection 3 & 0.130 & 0.128 \\
Selection 4 & 0.098 & 0.095 \\
Binary & 0.105 & 0.102 \\
Regression & 0.076 & 0.071 \\
Truth & 0.104 & 0.101 \\
\noalign{\smallskip}\hline
\end{tabular}
\end{table}
\end{center}

Both approaches, analytical with binary classification and regression, show an improvement with respect to the random solution choice, as expected.
An interesting observation emerges from the comparison of the fit performance employing different reconstruction techniques with the truth information. 
While the binary classifier very closely approaches the truth performance, the regression model even outperforms the fit using the truth distributions. What we might have considered as a {\em failure} of the DNN in reconstructing the truth angular distribution seems to encode some further important information for the distinction of the longitudinal polarization from the background, which might not be simply extracted from the truth angular distribution with respect to only one variable. 

We summarize the performance of the different methods exploited in this paper in Tables \ref{tab:selrmse} and \ref{tab:NNrmse2} where RMSE between the true and reconstructed $\cos\vartheta$ is used as a general metrics. DNN technique gives smaller values with respect to classic selection criteria, confirming the potential of the method according to different vector boson scattering processes.

\begin{center}
 \begin{table}
\caption{RMSE between the the true and reconstructed $\cos\vartheta$, calculated as a quantitative parameter to evaluate the performance of applied selection-criteria and DNN techniques in the semi-leptonic channel. 
We evaluate RMSE on samples containing a purely longitudinally polarized ($\mathrm{W}_{\text{L}}$) leptonically decaying $\mathrm{W}$ boson, a purely transversely polarized ($\mathrm{W}_{\text{T}}$) boson and an unpolarized $\mathrm{W}$ boson generated in the OSP {\tt PHANTOM} framework (Un. OSP). RMSE is equal to $0$ in the case of truth.}
\label{tab:selrmse}       
\begin{tabular}{llll}
\hline\noalign{\smallskip}
 & $\mathrm{W}_{\text{L}}$ & $\mathrm{W}_{\text{T}}$ & Un. OSP \\
\noalign{\smallskip}\hline\noalign{\smallskip}
Truth & $0.0$ & $0.0$ & $0.0$ \\
Random & $0.440$ & $0.347$ & $0.359$ \\
Selection 1 & $0.420$ & $0.303$ & $0.328$ \\
Selection 2 & $0.401$ & $0.269$ & $0.305$ \\
Selection 3 & $0.390$ & $0.274$ & $0.305$ \\
Selection 4 & $0.370$ & $0.228$ & $0.275$ \\
Binary & $0.357$ & $0.241$ & $0.270$ \\
Regression & $0.277$ & $0.200$ & $0.217$ \\
\noalign{\smallskip}\hline
\end{tabular}
\end{table}
\end{center}
\begin{center}
 \begin{table}
\caption{RMSE between the the true and reconstructed $\cos\vartheta$ calculated as a quantitative parameter to evaluate the performance of applied DNN techniques. We are comparing neural networks trained in the fully-leptonic channel.
Models are trained and evaluated on samples of purely longitudinally polarized ($\mathrm{W}_{\text{L}}$) and a purely transversely polarized ($\mathrm{W}_{\text{T}}$) leptonically decaying $\mathrm{W}$ bosons and then direct and indirect approaches are used, with an optional inclusion of the MAOS and MT2 variables in the training, as described in the text.
RMSE is equal to $0$ in the case of truth.}
\label{tab:NNrmse2}       
\begin{tabular}{llll}
\hline\noalign{\smallskip}
	DNN type &  $\mathrm{W}_{\text{L}}$ & $\mathrm{W}_{\text{T}}$ \\
\noalign{\smallskip}\hline\noalign{\smallskip}
direct & $0.291$ & $0.262$ \\
direct+MAOS & $0.291$ & $0.261$\\
indirect & $0.528$ & $0.605$\\
indirect+MAOS & $0.436$ & $0.493$\\
\noalign{\smallskip}\hline
\end{tabular}
\end{table}
\end{center}

\section{Conclusion}
\label{sec:conclusion}

Several approaches, in the direction of optimally reconstructing the full kinematics of $\mathrm{W}$ boson leptonic decays for the case of same-sign $\mathrm{WW}$ scattering with semi-leptonic and fully-leptonic decay channels have been presented in this paper, aiming at the extraction of the longitudinal polarization fraction of the $\mathrm{W}$ bosons in the physics process.  The binary classification and regression ML (DNN) approaches to reconstruct the longitudinal neutrino momentum have been proved to give a superior performance, if compared to more traditional approaches, even as advanced ones as MAOS.  The ability of the DNN approach to fully reconstruct the neutrino kinematics may be even more powerful in the case of resonant processes, where some intermediate heavy particle (e.g. the Higgs boson or some new, yet unknown, resonance)  decays into a  pair of $\mathrm{W}$ bosons and eventually into a final state involving neutrinos.  The full kinematic reconstruction of the final state particle kinematics allows the subsequent use of mass constraints for event selection in the case of SM resonance or invariant mass fitting in the case of BSM resonance, thus improving the signal-to-background separation and the discovery potential of new physics sear\-ches.

\begin{acknowledgements}
The authors would like to acknowledge the contribution of the COST Action CA16108 which initiated this work. Moreover, this work was supported by several STSM Grants from the COST Action CA16108. The authors are also indebted to the IBM System, and in particular to L. Laderchi, for his assistance
with the AC922 machine.
\end{acknowledgements}

\begin{appendix}
\section{PHANTOM generator (in a nutshell)}
\label{sec:app1}

\texttt{Phantom} is an event generator which is capable of simulating any set of diagrams with six partons in the final state
 at pp colliders at order $O(\alpha_{\text{EW}}^6 )$ and $O(\alpha_{\text{EW}}^4 \alpha_s^2 )$
including possible interferences between the two sets of diagrams. This includes all purely electroweak contributions as well as all contributions 
with one virtual or two external gluons. In the semi-leptonic channel we define the channel \texttt{mu vm}\_ whereas for full leptonic case the channel is defined in the configuration card with this parsing syntax \texttt{mu}\_ \texttt{e}\_ \texttt{vm ve}\footnote{Where the \_ stands for antiparticle}.
In this generation we removed all $b-$quarks and $t-$quarks contribution.
Phantom performs an exact calculation of the matrix-element at tree level, without using any production times decay approach for the W bosons interaction.
 Moreover it is computationally efficient since it is a dedicated generator for six fermions final states.
Phantom implements a method called On Shell Projection (OSP) to compute the amplitude of resonant contributions projecting
 (in the numerator) the four momenta of the decay particles on shell, as shown in Eq.~\ref{Eq:osp}.
 The formula can be seen as an on shell production times decay modulated by the Breit-Wigner width shape with all exact spin correlations.

\begin{equation}
A_f = \sum_{\lambda} \frac{A^{\mu}_{p,RES}(p,k)\epsilon_{\mu}^{\lambda}\epsilon_{\nu}^{\lambda*}A_d^{\nu}(k,q)}{k^2-M_W^2 +i\Gamma_WM_W} + A_{NON RES} 
\end{equation}

\begin{equation}
\label{Eq:osp}
\to \sum_{\lambda} \frac{A^{\mu}_{p,RES}(p,k_{OSP})\epsilon_{\mu,OSP}^{\lambda}\epsilon_{\nu,OSP}^{\lambda*}A_d^{\nu}(k_{OSP},q_{OSP})}{k^2-M_W^2 +i\Gamma_WM_W}
\end{equation}
\noindent
With this method, \texttt{PHANTOM} conserves the total four-mo\-men\-tum of the $\mathrm{WW}$ system, the direction of the two $\mathrm{W}$ bosons in the $\mathrm{WW}$ center-of-mass frame
and the direction of each charged lepton in his $\mathrm{W}$ center-of-mass frame.
In the case of semi leptonic single polarization/resonant study, we select the polarization for only one boson
whereas the other boson is kept unpolarized.

\section{Event generation configuration}
\label{sec:app2}

\texttt{PHANTOM} events used in this study have been produced with program version 1.6 
by configuring the generator according to the following parameters and sets of cuts, 
in particular cuts on forward/backward jets, missing $p_{\text{T}}$ and lepton masses to ensure the validity of generator polarization definition and selection.

\begin{itemize}
\item Parton distribution function: \texttt{NNPDF30\_nnlo\_as\_0118}.
\item Calculation type: $\alpha_{\text{EW}}^6$ at $13$ TeV CM energy.
\item Scale choice: (invariant mass of the 2 central jets and of 2 leptons)$ / \sqrt2$.
\end{itemize}

\begin{itemize}
\item $p_{\mathrm{T}}^{\ell} > 20$ GeV;
\item $|\eta^{\ell}| < 2.5$;
\item $p_{\mathrm{T}}^{\mathrm{miss}} > 40$ GeV;
\item $p_{\mathrm{T}}^{j} > 30$ GeV;
\item $|\eta^{j}| < 4.5$;
\item $|\Delta\eta_{jj}| > 2.5$;
\item $m_{jj} > 500$ GeV;
\item $\Delta R_{\ell\ell} > 0.3$;
\item $\Delta R_{j\ell} > 0.3$;
\end{itemize}

\section{MAOS solution}
\label{sec:app_maos}

Equality $m^{(1)}_{\text{T}}(\vec{p}_1) = m^{(2)}_{\text{T}}(\vec{p}_{\text{T}} - \vec{p}_1)$ defines a curve in the phase space of the first neutrino trial momentum $\vec{p}_1$. 
This curve equation can be expressed analytically in the polar coordinates $(p_1,\varphi)$. Let us introduce the parameters:

\begin{eqnarray}
\varphi_0 &=& \arccos\left(\frac{\vec{p}_{\text{T}}^{\,\ell\ell}\cdot\vec{p}_{\text{T}}^{\,\text{miss}}}{|\vec{p}_{\text{T}}^{\,\ell\ell}||\vec{p}_{\text{T}}^{\,\text{miss}}|}\right),\notag\\
c &=& (\vec{p}_{\text{T}}^{\,e}\cdot \vec{p}_{\text{T}}^{\,\text{miss}})^2 - |\vec{p}_{\text{T}}^{\,\text{miss}}|^2|\vec{p}_{\text{T}}^{\,e}|^2,\notag
\end{eqnarray}

where $\vec{p}_{\text{T}}^{\,\ell\ell} = \vec{p}_{\text{T}}^{\,\mu} + \vec{p}_{\text{T}}^{\,e}$. The intersection curve can be written as:

\begin{eqnarray*}
\begin{aligned}
f(\varphi) = \left(|\vec{p}_{\text{T}}^{\,\mu}|^2 + |\vec{p}_{\text{T}}^{\,\ell\ell}|^2\right.&\left.\cos^2\varphi - 2|\vec{p}_{\text{T}}^{\,\mu}||\vec{p}_{\text{T}}^{\,\ell\ell}|\cos\varphi - |\vec{p}_{\text{T}}^{\,e}|^2\right),\\
g(\varphi) = \left(2(|\vec{p}_{\text{T}}^{\,\mu}| - |\vec{p}_{\text{T}}^{\,\ell\ell}|\right.&\left. \cos\varphi)\vec{p}_{\text{T}}^{\,e}\cdot \vec{p}_{\text{T}}^{\,\text{miss}}\right.\\ &\left.+ 2|\vec{p}_{\text{T}}^{\,e}|^2|\vec{p}_{\text{T}}^{\,\text{miss}}|\cos(\varphi + \varphi_0)\right),
\end{aligned}\\
p_1 = \frac{-g(\varphi) \pm \sqrt{g(\varphi)^2 - 4cf(\varphi)}}{2f(\varphi)}.\qquad\qquad\qquad\qquad\,\,\,\,\,\,\,
\end{eqnarray*}

$m_{\text{T2}}$ can be searched numerically by defining an arbitrary number of equidistant points in $\varphi$ and searching for a minimal value of $m^{(1)}_{\text{T}}(\vec{p}_1(\varphi))$. This point defines $\vec{p}^{\,\nu_{\mu}}_{\text{T}}$, $\vec{p}^{\,\nu_e}_{\text{T}}$ can be obtained from the minimization bond as $\vec{p}^{\,\nu_e}_{\text{T}} = \vec{p}_{\text{T}}^{\,\text{miss}} - \vec{p}^{\,\nu_{\mu}}_{\text{T}}$.

\section{Neural network details}
\label{sec:nndet}
In this work, the most basic type of neural network has been adopted: a {\em feed forward neural network}.
In a feed forward neural network the nodes are organized in layers. The input variables $(x_1 , ..., x_n )$ are placed in the input layer.
The training variables used in this work are related to lepton part and hadrons: 
\begin{itemize}
\item $ \mu$: $E,p_x,p_y,p_z, p_t,\eta,\phi$,
\item $\nu_{\mu}$:$p_x,p_y$ 
\item $q_1$: $E,p_x,p_y,p_z$
\item $q_2$:,$E,p_x,p_y,p_z$, 
\item $q_3$:,$E,p_x,p_y,p_z$,  
\item $q_4$:,$E,p_x,p_y,p_z$, 
\item $event0\_mass,event1\_mass$
\end{itemize}

where $q_i$ refers to outgoing quarks and $event0\_mass$, \\
$event1\_mass$ refer to the fact that we have two pos\-si\-bi\-li\-ties for the neutrino $z$-component of momenta and energy, 
eventually leading to a double measurable invariant masses of the event.

 Each value of the input layer is passed to each node of the next layer with the corresponding weight. The output is then passed to the next layer and so on,
  where intermediate layers are called {\em hidden layers}.
Finally, the last output layer returns the results of the computation of the neural network.

In this study, {\em deep} neural networks have been used: this adjective is used to categorize neural networks with a lot of nodes for each layer 
$(10^1,10^3)$ and up to ten hidden layers.
Increasing the number of parameters can help the neural network to accomplish more difficult classification tasks because its approximation power increases \cite{Lista:2017jsy}.

Before starting the training procedure, a Gaussian standardization technique is applied to the whole training dataset, signal and background:
 each variable is centered in 0 and scaled to unit variance independently. This procedures helps the neural network training because reduces the
  scale differences of its inputs.

\begin{figure*}
	\centering
	\includegraphics[width=0.40\textwidth]{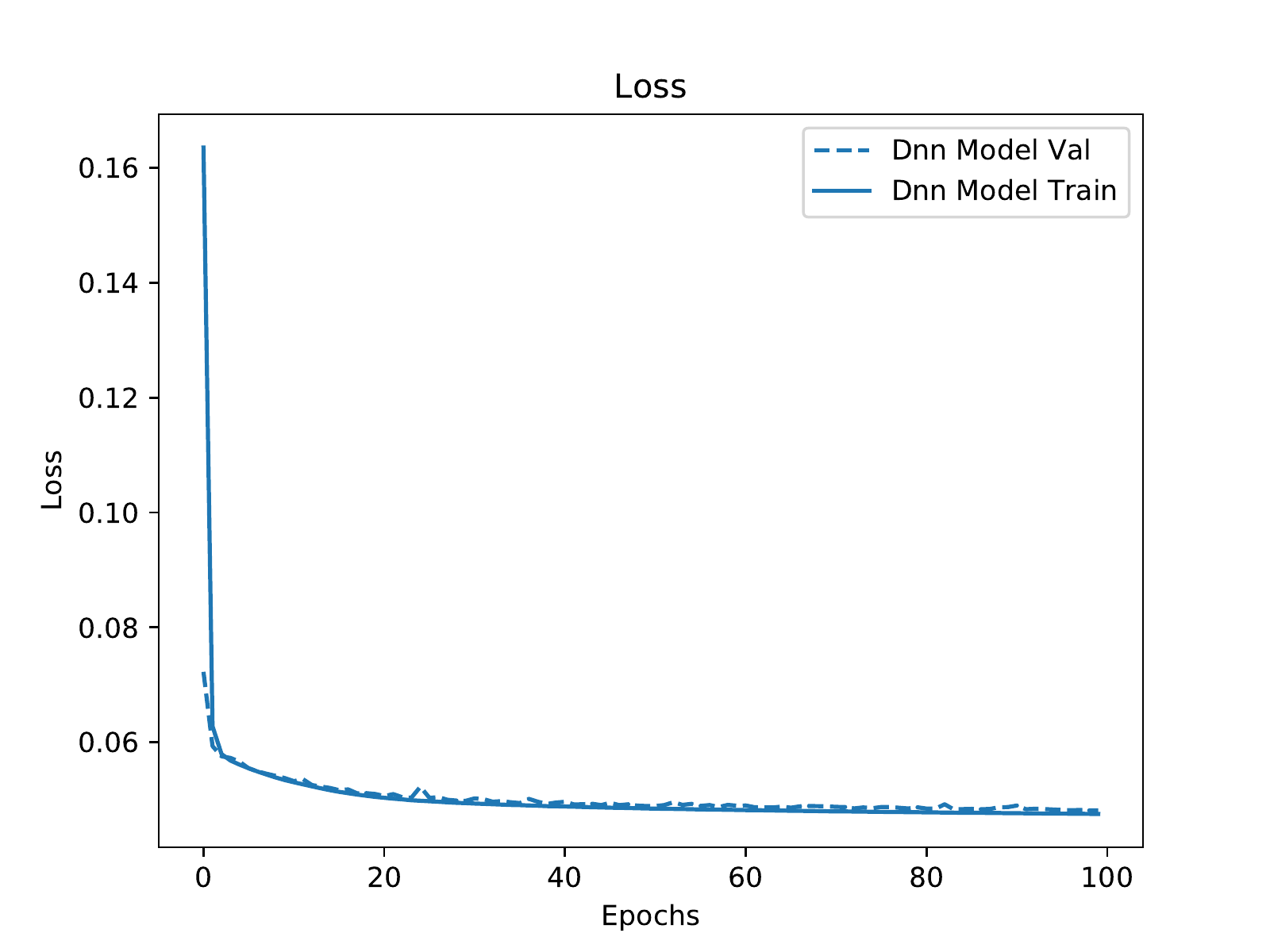}
	\includegraphics[width=0.40\textwidth]{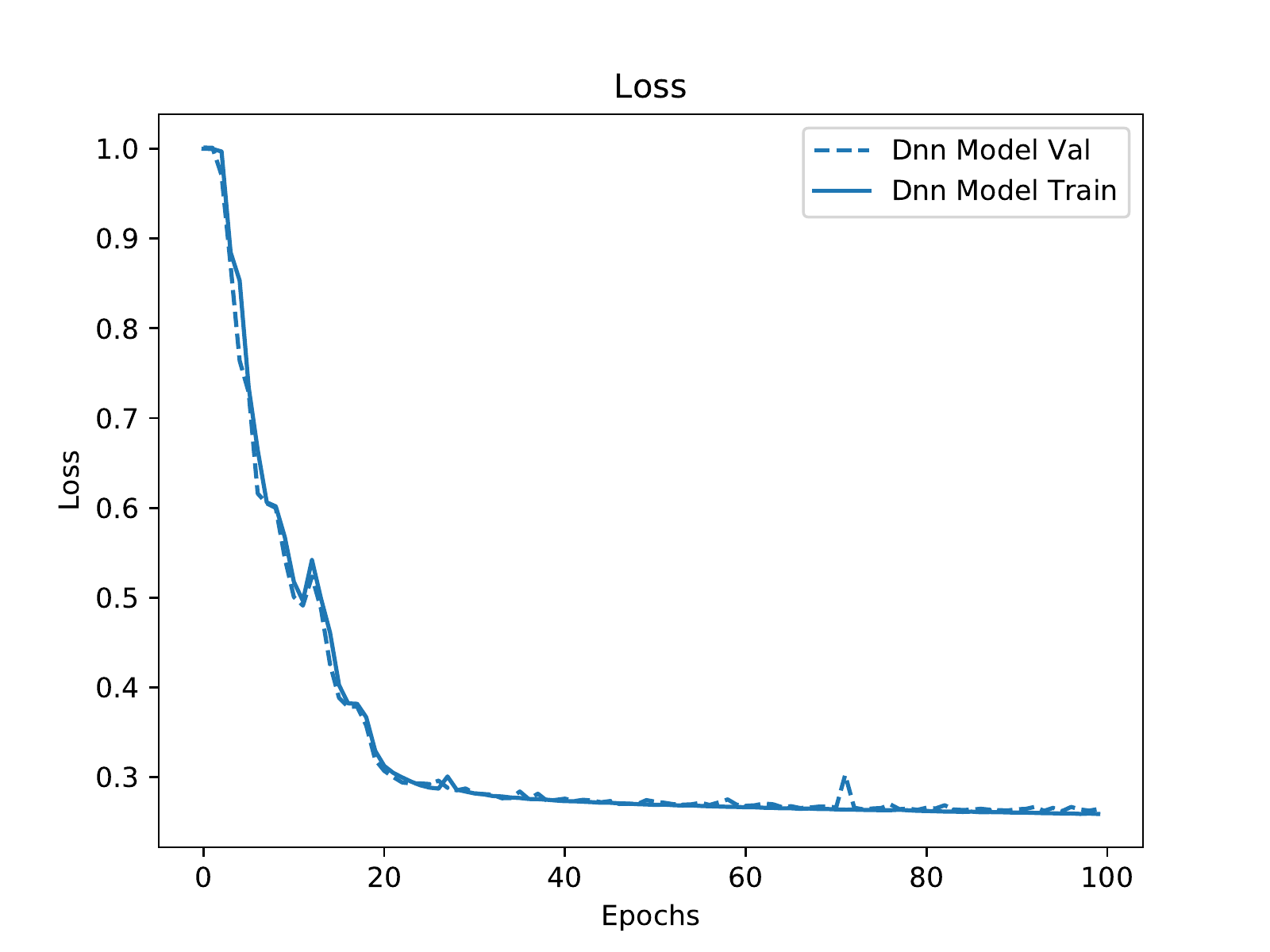}
	\caption{Shape of the loss function related to a specific neural network topology used for regression in the semi-leptonic and fully-leptonic channel. The decreasing of loss values a minimum value can be appreciated so that represents a good convergence of the neural network parameters according to the number of hidden layers, neurons and batch-size. }
	\label{fig:NNgood}
\end{figure*}

\begin{figure*}
	\centering
	\includegraphics[width=0.40\textwidth]{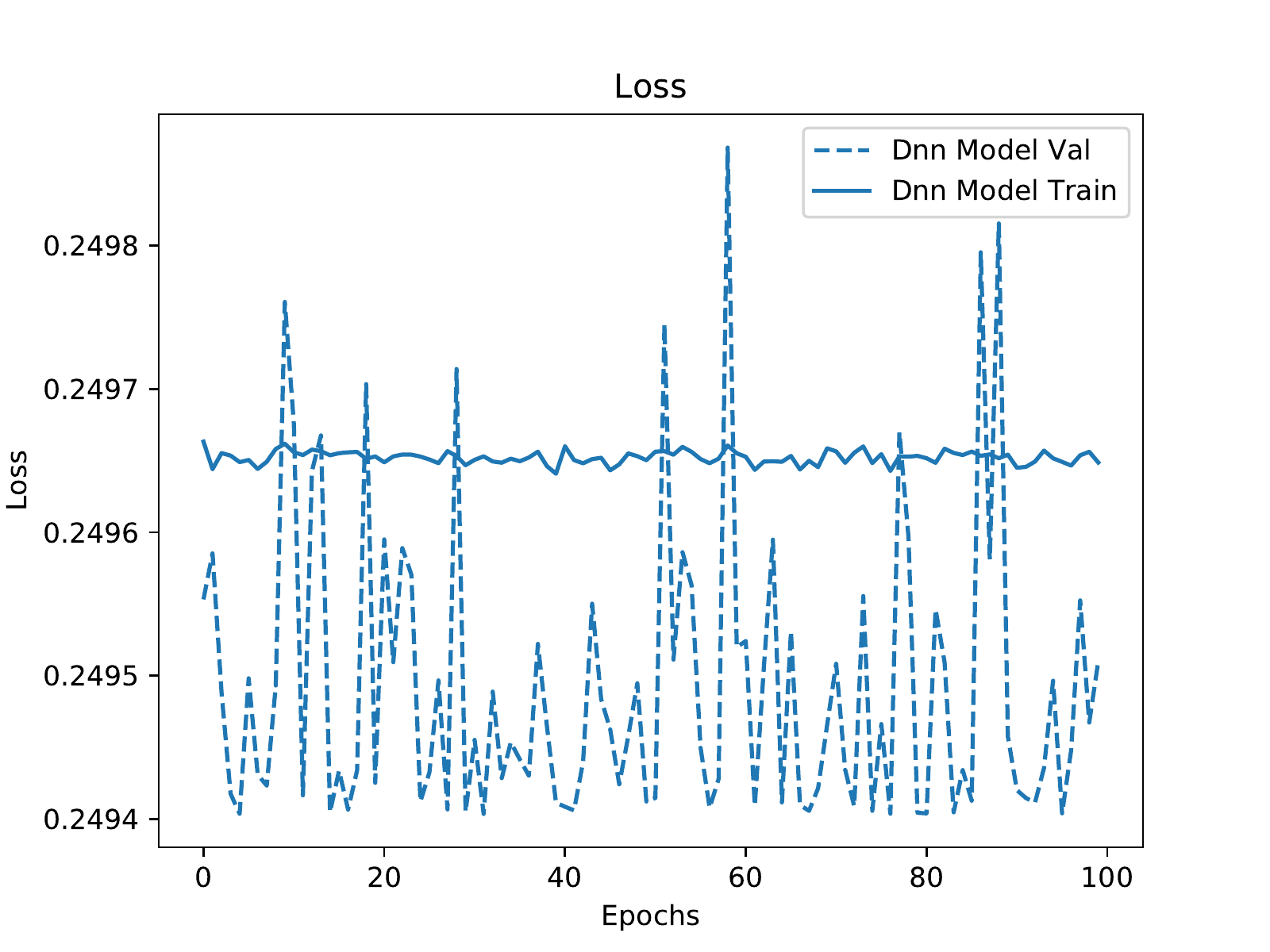}
	\includegraphics[width=0.40\textwidth]{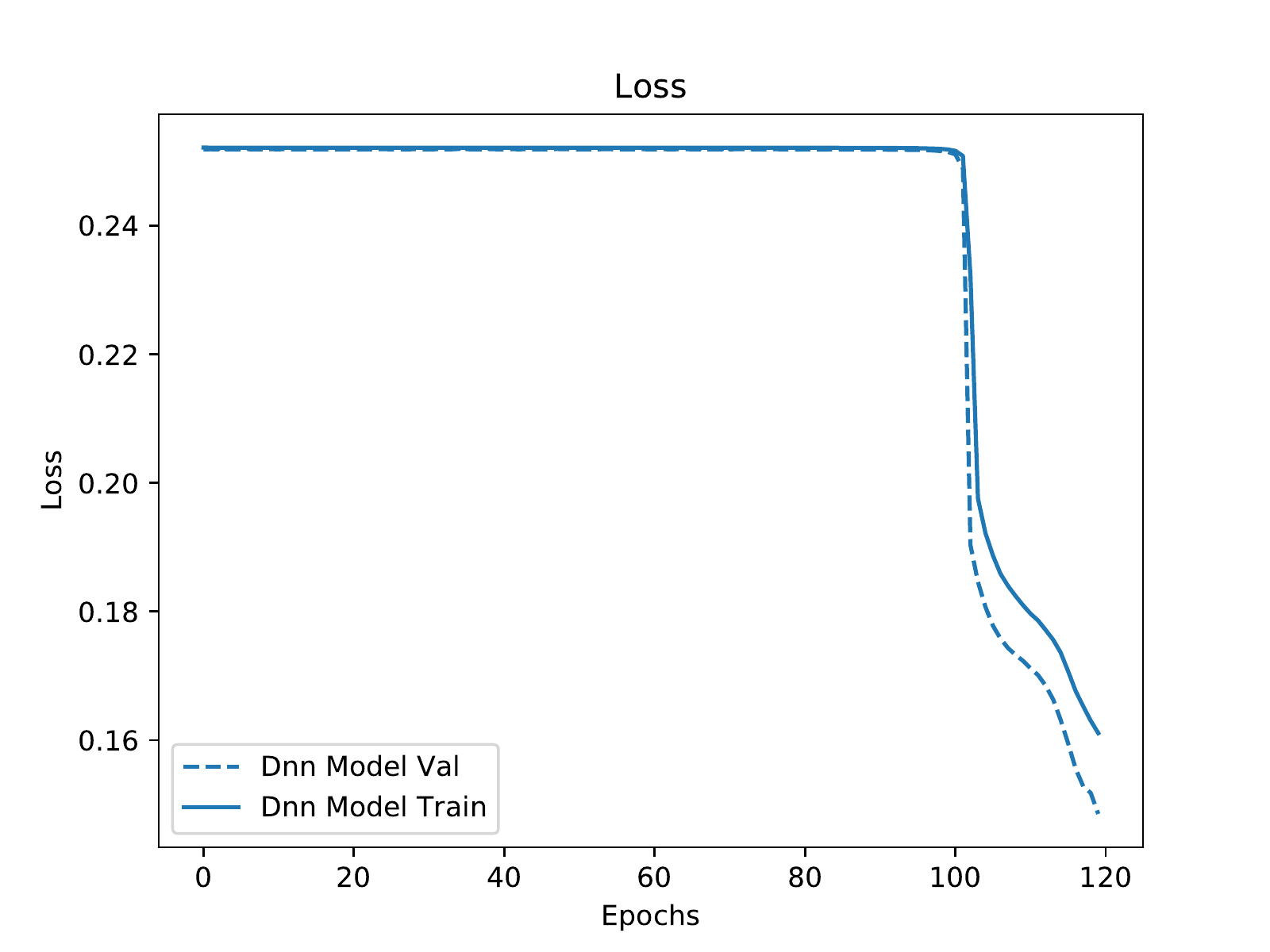}
	\caption{Shape of the loss function related to a specific neural network topology used for regression in the semi-leptonic and full-leptonic channel. 
	In this case the loss function is not converging, in particular in the first case the network is not improving its parameters at all giving a flat loss function and relatively unstable results for the validation dataset.
	In the second case the neural networking is not improving and only after a relatively high number of epochs the loss is starting to evolve. For the optimal treatment of each particular convergence scenario an early stop automation can be employed so that the number of training epochs in each case can be tuned according to the learning rate and loss function behavior. }
	\label{fig:NNbad}
\end{figure*}

For sake of completeness we include representative plots of loss function behavior in Figures \ref{fig:NNgood} and \ref{fig:NNbad} for different DNN scenarios and configurations in order to highlight the performance of our framework.

\begin{center}
 \begin{table}
\caption{Number of events used for each sample in the training, validation and test datasets.}
\label{tab:NNdataset}       
\begin{tabular}{llll}
\hline\noalign{\smallskip}
sample & training & validation & test  \\
\noalign{\smallskip}\hline\noalign{\smallskip}
unpolarized & $530219$ & $530219$ & $706959$\\
longitudinal & $264052$ & $264052$ & $352070$\\
transverse & $269769$ & $269769$ & $359692$\\
full-computation & $535236$ & $535236$ & $713648$\\
\noalign{\smallskip}\hline
\end{tabular}
\end{table}
\end{center}

\section{Computation details}
All activities have been performed in two main steps.
First all the events were generated on a PC computing cluster using the HTCondorCE
that allows the user to parallelize all the computation splitting it into several nodes.

Secondly, once all the events were created, we converted them into Root files \cite{citeulike:root} to perform common data preparation 
and finally we converted them into pandas arrays to get train, test and validation dataset.
These data was migrated to an IBM Power9 machine to perform all the training and evaluation steps.
IBM Power System AC922 is widely used in general context for its performance for analytics, artificial intelligence (AI), and modern HPC. 
The Power AC922 is engineered to be the most powerful training platform available, providing the data and compute-intensive infrastructure
 needed to deliver faster time to insights. Data scientists get to use their favorite tools without sacrificing speed and performance, 
 while IT leaders get the proven infrastructure to accelerate time to value. This is IT infrastructure redesigned for enterprise AI\footnote{\url{https://www.ibm.com/us-en/marketplace/power-systems-ac922}}.

\end{appendix}

\bibliographystyle{spphys}       
\bibliography{DNN_neutrino_paper}   

%
%

\end{document}